\documentclass[acmsmall]{acmart}
\AtBeginDocument{%
  }

\setcopyright{cc}
\setcctype{by}
\acmJournal{PACMHCI}
\acmYear{2026} \acmVolume{10} \acmNumber{6} \acmArticle{CSCW152}
\acmMonth{10} \acmDOI{10.1145/3817000}

\usepackage{float}
\usepackage{array}
\usepackage{booktabs}

\begin{document}

\title{Silence and Noise: Self-censorship and Opinion Expression on Social Media}

\author{Xinyu Wang}
\email{xzw5184@psu.edu}
\affiliation{%
  \institution{The Pennsylvania State University}
  \country{USA}
}

\author{Emma Carpenetti}
\affiliation{%
  \institution{The Pennsylvania State University}
  \country{USA}
}
\email{erc5513@psu.edu}

\author{Bruce Desmarais}
\affiliation{%
  \institution{The Pennsylvania State University}
  \country{USA}
}
\email{bdesmarais@psu.edu}

\author{Sarah Rajtmajer}
\affiliation{%
  \institution{The Pennsylvania State University}
  \country{USA}
}
\email{smr48@psu.edu}

\renewcommand{\shortauthors}{Wang et al.}

\begin{abstract}
Unlike the more observable phenomenon of group opinion reinforcement, self-censorship online has received comparatively less attention. Our goal in this work is to dissect the phenomena of self-censorship and to examine the implications of restrained expression for participation in public discourse, particularly in polarized contexts. We explore how social media users express their opinions online through analyses of 390 survey responses and 20 semi-structured interviews using a mixed-methods approach. We ask social media users about the differences between their publicly shared opinions and privately held beliefs, highlighting the influence of contextual factors on self-expression.  
Our findings show that self-censorship is associated with community context; social media users embedded within larger audiences, with lower posting frequency and perceived support, are less likely to express their opinions, and those who do speak often adjust their expressed views to align with perceived group norms.
The study complements the rich literature on echo chambers and opinion reinforcement on social media platforms, highlighting the silence within the noise and its potential consequences for public discourse, which have become increasingly pertinent in an era where online platforms are pivotal to social and political narratives.
\end{abstract}



\received{May 13, 2025}

\received[revised]{January 13, 2026}

\received[accepted]{April 9, 2026}

\maketitle
\section{Introduction}
Social media encourages the establishment of communities which frame and reinforce a shared narrative (see, e.g., \cite{cinelli2021echo}). 
In recent years, social media studies have explored opinion formation, e.g., through the lens of echo chambers and filter bubbles \cite{pennycook2018prior,unkelbach2021mere}, and a variety of models have been invoked to describe these phenomena \cite{shang2019resilient}. One overarching concern is that the algorithms behind content dissemination drive polarization and fail to mitigate the spread of misinformation. The dynamics of misinformation and polarized content have been a topic of significant focus in recent years  \cite{allcott2019trends,zimdars2020fake, karlova2013social,kumar2014detecting,del2016spreading,shin2018diffusion,vosoughi2018spread}. 

Work in sociology, psychology, and communication has long emphasized the distinction between privately held opinions and publicly expressed views, particularly in social environments characterized by conformity pressures \cite{jack2011reflections, hampton2014social,gastner2018consensus}. This distinction underlies a rich theoretical tradition documenting how individuals remain silent or adjust their expressed opinions in response to perceived social pressure and anticipated evaluation \cite{asch1951effects, crutchfield1955conformity, gearhart2015something, gibson2023keeping}. For example, the Spiral of Silence theory \cite{noelle1974spiral} posits that the fear of isolation drives individuals to monitor their environment; if they perceive that their view is in the minority, they are less likely to voice it \cite{gibson2023keeping}. 

While these theories originated in the context of mass media, their application to the Computer-Supported Cooperative Work (CSCW) domain introduces new complexities. Digital environments introduce unique ``affordances'', such as context collapse \cite{marwick2011tweet}, that fundamentally alter how individuals navigate the tension between their private identity and their public persona \cite{hollenbaugh2021self}. Recent empirical work has documented manifestations of self-censorship online. Hampton et al. \cite{hampton2014social} found evidence of spiral-of-silence effects on social media; Gera et al. \cite{gera2020hesitation} showed that willingness to express opinions online varies by topic, with hesitation commonly driven by anticipated audience reactions; Warner and Wang \cite{warner2019self} emphasized the role of privacy concerns and information control in shaping self-censorship on social networking sites.

However, several gaps remain in our understanding of self-censorship in online contexts. (1) There is a lack of research into the gradient of self-censorship, specifically the distinction between complete silence and opinion conformity. Self-silencing theory suggests that these behaviors exist on a continuum but prior work has not examined how contextual factors are associated with different points along this continuum. (2) It remains unclear how contextual factors are associated with users' corresponding expression decisions.
(3) We lack understanding of users' own perceptions of the consequences of restrained and reinforced expression and what interventions they believe might support healthier discourse. 
This paper addresses these gaps by adopting a mixed-methods approach to quantify the divergence between public and private opinions. We move beyond the ``observed post'' as the unit of analysis to examine the deliberative process users undergo before deciding to speak or remain silent. Methodologically, we integrate survey data (N=390) that captures specific differences between private beliefs and public expression across six topical domains, with semi-structured interviews (N=20) that explore the reasoning behind these patterns. 

Our work is scaffolded by the following research questions: 

\vspace{0.1cm}

\noindent\textbf{RQ1}: Is there a discrepancy between the opinions individuals share publicly through social media (we will call these \emph{public opinions}) and their privately held beliefs (we will call these \emph{private opinions})?

\noindent\textbf{RQ2}: How are contextual factors, e.g., topic, perceived community, and posting behavior, associated with users’ willingness to share opinions and differences between private and public expression on social media?

\noindent\textbf{RQ3}: What are users' perceived consequences of opinion conformity and reinforcement, and what interventions can be implemented to promote healthier social media ecosystems?

\vspace{0.1cm}

\noindent By way of these questions, our goal is to explore how self-censorship--literally ``the act of preventing oneself from speaking'' \cite{das2013self}--contributes to online conversation dynamics. Our findings suggest that self-censorship and milder versions of self-censorship, including posting subtle opinions and engaging passively, can lead to reduced diversity of viewpoints 
and inadvertently encourage more extreme views and misinformation to dominate public discourse. 

We examine how users navigate these dynamics and identify design directions that can better support expressive confidence online. These include: re-imagining fact-checking to support credible expression; developing incentive structures that elevate respectful and inclusive interactions; and strengthening accountability through both transparent internal moderation and external regulatory oversight. By emphasizing collaborative responsibility across platforms, user communities, and regulators, we aim to advance a vision for social media environments where individuals are not only protected from harm, but also actively encouraged to participate in meaningful and open public discourse.

\section{Related Work}

\subsection{Theoretical Motivations}

Our study is grounded in a set of socio-psychological theories that explain how individuals form, suppress, and express opinions in social contexts, particularly within online environments. We use them to frame concrete behaviors that can be observed through self-reported discrepancies between private beliefs and public expression.

\noindent \textbf{Self-silencing theory.} Self-silencing theory provides a framework for understanding restrained opinion expression as a set of related outcomes shaped by social pressure and anticipated evaluation. Within this broader landscape of social influence, self-silencing refers to the inclination to restrain self-expression and action in order to protect social ties or avert anticipated retaliation, for example, by suppressing one's true opinions to preserve a positive public image \cite{jack2011reflections}. Prior research documents a “spiral of silence” in online contexts, where individuals refrain from speaking publicly about polarizing issues when they perceive their views as misaligned with prevailing opinions \cite{hampton2014social,gearhart2015something,sohn2022spiral,chaudhry2020expressing}. In this paper, we use the term \emph{self-censorship} to refer to these forms of restrained expression, encompassing both complete silence and opinion conformity. A common response is the decision to remain silent, where individuals choose not to express their views at all in response to perceived social risk. Importantly, theories of social influence also describe situations in which expression is not withheld but adjusted. Individuals may continue to participate while modifying the intensity of their expressed views, leading to discrepancies between private beliefs and public expression. On the other hand, opinion reinforcement refers to behaviors in which publicly expressed opinions differ in intensity from an individual’s original position, often becoming more positive or negative in sentiment \cite{lim2022opinion}. Reinforcement and self-censorship can be understood as alternative responses that are discussed within the same theoretical landscape of social evaluation and visibility, rather than as mutually exclusive outcomes.

\noindent \textbf{Social identity theory.} Opinion expression on social media is inherently shaped by audience context. While self-silencing theory is used to define the forms of restrained expression we examine, social identity theory provides a lens for understanding the contextual factors associated with these outcomes. Social identity theory suggests that individuals tend to understand themselves and others in relation to social groups, and that group membership provides a powerful lens through which opinions are interpreted and evaluated \cite{tajfel1979integrative,tajfel2004social,stets2000identity}. When a group identity becomes prominent, individuals develop their self-concept around paradigmatic ingroup characteristics \cite{brewer1999psychology}. They are motivated to differentiate between their ingroup and relevant outgroups \cite{li2020real}. In online social communities, these dynamics shape how users anticipate audience reactions and evaluate the social costs of expression, influencing their willingness to share opinions \cite{greene2004social,trepte2017social}. Classic experimental and contemporary work further shows that perceived social approval and disapproval can distort expressed opinions even when private beliefs remain unchanged \cite{asch1951effects,crutchfield1955conformity,abbink2017peer}. Related phenomena such as pluralistic ignorance \cite{allport1924social} and the enforcement of unpopular norms \cite{centola2005emperor,willer2009false} illustrate how misperceptions of group consensus can further constrain public expression. This perspective motivates the inclusion of perceived support as a key variable, capturing the extent to which individuals perceive alignment with their audience, as well as community size, which reflects the scale and heterogeneity of the audience in which these identity-based evaluations take place. Prior work has shown that anticipated acceptance or rejection within one’s social group is associated with decisions to withhold or modify expression \cite{burnett2022self, bar2017self}, while evidence on the role of audience size in shaping self-censorship remains mixed \cite{kwon2015unspeaking, coleman1988social, brandtzaeg2010too, das2013self}.

\noindent \textbf{Self-presentation theory.} Self-presentation is the behavior of conveying information about oneself, driven by the motivation to align one's self-display with the expectations and preferences of the audience \cite{baumeister1987self}. In this study, we draw on self-presentation theory to guide the qualitative analysis of how individuals manage expression in social media contexts. This perspective helps interpret patterns of expression management in relation to anticipated social evaluation.
Originally, theories of self-presentation were centered on in-person interactions, exploring how individuals perform and project their self-image in face-to-face settings \cite{goffman1949presentation,schlenker1985identity}. However, as digital media, particularly social media, have become increasingly integral to daily life, these traditional theories have evolved and now encompass the nuances of self-presentation in digital formats, adapting to the dynamics of online communication and identity expression \cite{hollenbaugh2021self}. The anchoring of an individual's online identity in their offline identity and physical space constrains the extent of highly controlled and selective self-presentation online, as the varying degrees of anonymity provided by certain social media platforms influence these dynamics \cite{hollenbaugh2021self}. Within this framework, discrepancies between private beliefs and public expression can be understood as strategic adjustments that allow individuals to maintain a favorable public image while managing potential social costs. 
In addition to self-initiated factors, the environment in which individuals interact plays a significant role in shaping their online behavior. Platform affordances, such as anonymity and the differing interaction dynamics across social media platforms, influence self-presentation and the willingness of self-expression \cite{oz2024platform}. These contextual conditions also shape how individuals interpret topic sensitivity; prior work shows that perceived topic polarity is associated with variation in willingness to express opinions, with more controversial topics corresponding to higher levels of hesitation and self-censorship \cite{zhao2025mapping, filak2009expanding}.
Finally, behavioral engagement provides an additional lens for understanding expression dynamics. Posting frequency reflects how often individuals participate in opinion expression and has been linked to audience-related factors such as community size \cite{racca2018relating}.

Rather than taking these theories as competing explanations, we suggest that they offer complementary lenses for interpreting observable patterns in participants' reported willingness to express opinions on social media. Self-silencing theory defines the forms of restrained expression we examine (complete silence vs. adjusted expression). Social identity theory helps explain when these behaviors occur by highlighting the role of group context and perceived alignment. Self-presentation theory guides our qualitative understanding of how individuals reason about and manage these tensions. Table~\ref{tab:theory} summarizes how these theoretical perspectives inform the selection of key variables used in our subsequent analyses.

\begin{table}[H]
\tiny
\centering
\begin{tabular}{p{1.2cm}p{2.3cm}p{9.2cm}}
\hline
Variable Type & Variable &  Theoretical Grounding \\
\hline
Dependent & Decision to remain silent &
This variable captures complete self-censorship, defined as the choice not to express an opinion publicly. Self-silencing theory and the spiral of silence framework suggest that individuals are less likely to voice their views when they perceive their opinion to be in the minority or anticipate negative social evaluation \cite{hampton2014social, gearhart2015something, sohn2022spiral, chaudhry2020expressing}. This outcome represents one endpoint of the continuum of restrained expression examined in this study. \\

Dependent & Discrepancy between private and public opinions &
This variable captures partial self-censorship in the form of opinion adjustment, where publicly expressed views differ from privately held beliefs. Self-silencing theory conceptualizes such adjustments as strategies to manage social risk while maintaining participation \cite{jack2011reflections}. In parallel, self-presentation theory suggests that individuals selectively modify expressed opinions to align with perceived audience expectations and maintain a favorable public image \cite{baumeister1987self, goffman1949presentation}.  \\

Independent & Perceived polarity  & Prior work shows that perceived topic polarity is associated with variation in opinion expression and restraint, with more controversial topics eliciting higher levels of hesitation and self-censorship \cite{zhao2025mapping, filak2009expanding}. Section~\ref{polvexp} further documents topic-dependent patterns of expression in this study. \\
Independent & Perceived support & This variable captures perceived social alignment within one’s audience. Work on self-silencing emphasizes that anticipated acceptance or rejection by others is associated with decisions to withhold or modify expression \cite{burnett2022self, bar2017self}.  \\
Independent & Community size & This variable captures the scale of the potential audience for expression and reflects structural aspects of audience context discussed in group identity theory \cite{tajfel1979integrative,tajfel2004social,stets2000identity}. Prior research offers mixed evidence on its relationship with self-censorship behaviors \cite{kwon2015unspeaking, coleman1988social, brandtzaeg2010too, das2013self}. \\
Mediator & Posting frequency  & Prior work suggests that audience scale is associated with how frequently individuals engage in posting behaviors \cite{racca2018relating}. \\
\hline
\end{tabular}
\caption{Model variables and theoretical rationale.}
\label{tab:theory}
\end{table}

\subsection{Opinion Dynamics on Social Media}

Prior literature has posited a divergence between individuals' publicly expressed positions and privately held beliefs \cite{king1981conflicts,rose2007going}. This divergence can manifest as either exaggeration or suppression of true opinions, influenced by social and individual factors \cite{konovalova2023social,neureiter2021trust}. 
Research on opinion dynamics in social media has emphasized the reinforcement of views through mechanisms such as echo chambers and filter bubbles \cite{barbera2020social, bruns2017echo, guess2018avoiding, rhodes2022filter}. These phenomena describe how users are increasingly exposed to homogeneous viewpoints that reinforce their existing beliefs, either due to network structure \cite{barbera2015tweeting, cinelli2021echo, garimella2018political} or algorithmic design \cite{nguyen2014exploring, bozdag2013bias}. These dynamics have often been associated with increased political polarization and the reinforcement of partisan attitudes \cite{bail2018exposure, hobolt2024polarizing, gillani2018me}. However, this body of literature centers almost exclusively on vocal participation—the conversations users choose to publicly share, amplify, or engage with—and often takes these conversations at face value through content analysis. Such approaches overlook the ways individuals withhold their true opinions, which can take the form of complete silence or the expression of a distorted version of their views. This self-censorship—what is not said or not explicitly said—also plays a critical role in shaping online opinion dynamics.

Recent studies have examined hesitation and restraint in online opinion expression. 
For example, Gera et al. \cite{gera2020hesitation} surveyed college students and found that posting willingness varies by topic, and these concerns are strong in politically diverse audiences, where anticipated social sanctions increase reluctance to post \cite{weeks2024too}. 
Gibson et al. \cite{gibson2023keeping} found that nearly half of Americans self-censor their opinions online, often in response to political polarization and fear of social repercussions.  Other studies have explored distinct underlying motivations. Warner et al. \cite{warner2019self}  emphasized privacy-related factors such as perceived vulnerability and information control as key drivers of self-censorship on social networking sites. Dubois et al. \cite{dubois2018self} further showed that belief in the political relevance of social media consistently predicts self-censorship across several countries, while impression management and exposure to hostile interaction have also been shown to encourage restraint in online political discussions \cite{powers2019shouting,juncosa2024toxic}. Howe et al. \cite{howe2023self} demonstrated that self-censorship can also serve as a strategy for limiting the spread of misinformation, showing that prompting users to share only content they believe is true significantly improves discernment.  Together, these studies suggest that self-censorship is a common and multifaceted behavior, shaped by a range of socio-psychological and contextual factors. Our work contributes to the understanding of self-censorship by examining how users navigate the tension between private beliefs and public expression, focusing on how perceived social context and group dynamics are associated with what individuals choose to share, and exploring the relationship between echo chambers and self-censorship.
\section{Research Methods}

\subsection{Survey Study}

We conducted a structured survey using Qualtrics, a platform widely utilized in academic research for survey design and deployment \cite{eyal2021data}. 

\subsubsection{Survey design.} The survey included Likert-scale items, slider-scale items, and open-ended questions. The survey was structured to measure contextual factors which prior literature has suggested may impact opinion-sharing on social media. Demographic and social media usage questions used in the study were adapted from existing work and slightly modified to suit the research objectives \cite{lee2021digital}. An outline of the survey follows here, and the full survey is provided in Appendix 1.

\noindent \textbf{Demographic information.}
We collected demographic data through multiple-choice questions on gender, age, race/ethnicity, education, and political identification, with a follow-up question for independents to specify a party leaning.

\noindent \textbf{Social media usage.}
In addition to general social media usage questions, we asked participants to share, for example, how often they post and the size of their audience. These items operationalize behavioral engagement and community size, which are central contextual dimensions in our theoretical framing and subsequent models.

\noindent \textbf{Opinion expression.}
We explored discrepancies between users' public and private opinions, prevalence of self-censorship, and habits around opinion sharing on social media. The statement-based items were designed to directly capture participants’ self-reported experiences of opinion restraint and adjustment. In addition, we specifically assessed participants' opinion-sharing behaviors with respect to six topics: politics, social justice, health, environment, technology, and religion. We asked participants to rate the polarity of each topic. The research team discussed and shortlisted six topical domains that are commonly encountered on social media and that differ in perceived polarity and public visibility. This selection provides a shared reference frame for participants while allowing examination of opinion expression across domains with varying levels of controversy.

\noindent \textbf{Topic-specific responses.}
We asked participants to share their private opinions on specific issues related to each of the six focal topics, e.g., climate change (see Appendix 1 for list of issues). For each topic, the research team selected a single, widely recognizable issue through majority agreement to serve as a shared reference point, anchoring responses in a concrete context and reducing ambiguity in interpretation. We measured their stance, perceived polarization, community support, and reported posting behaviors for each. 

\noindent \textbf{Open-ended questions.}
Participants reflected on discrepancies between their public posts and private beliefs with respect to our focal topics. We asked participants to discuss the frequency and impact of such discrepancies. We asked them to describe specific posts they have shared and compare those to their privately held beliefs. These questions were included to capture context and reasoning that may not be well represented by closed-ended items.

\subsubsection{Target sample size.} We calculated target sample size by setting the confidence level at 95\%, margin of error at 5\%, and using the estimated number of social media users in the U.S. in 2023 (302.35 million)\footnote{https://www.demandsage.com/social-media-users/} as the target population size. This calculation resulted in a required sample size of 385. Accordingly, we collected 390 responses through Prolific. 

\subsubsection{Recruitment and compensation.} Survey recruitment was managed through the crowd-sourcing platform Prolific, which supports fine-grained population sampling and has been shown to generate high-quality response data \cite{douglas2023data}.  For inclusion in the study, participants were required to be located in the United States, at least 18 years of age, and report using English as their primary language of communication on social media platforms. Participant demographics are provided in Table \ref{tab:statistics} (see Appendix 3). Each participant was paid \$4 upon approval, which meets the minimum wage standard at \$12/hr. In total, \$2133.33 was spent for survey participant compensation (including service fees).

\subsubsection{Quality checks.} We undertook two-stage verification for survey data. Responses were required to pass both an attention check \cite{oppenheimer2009instructional} and manual verification of Qualtrics-flagged potential bots. All 390 responses were concluded to be valid (see Appendix 3 for details about quality check procedure).

\subsubsection{Pilot study.} A pilot version of the survey was conducted with 10 participants from Prolific. 
They were asked to provide feedback on the clarity of question phrasing and the comprehensibility of question framing after completing each section. Based on their responses, questions were refined to enhance clarity. There is no overlap between these 10 and the 390 participants in the reported survey study. 
The pilot study also informed the selection of the final statements used to measure opinion difference, conformity, and reinforcement. For each construct, an initial pool of five candidate statements was included in the pilot survey to capture slightly different formulations of the same underlying concept. Responses from the pilot were reviewed to assess whether the statements were interpreted consistently, whether they elicited similar response patterns, and whether any statements caused confusion or redundancy. Based on this review, three statements per construct were selected for the main survey. These selected statements showed consistent interpretation across participants, aligned closely with the intended theoretical constructs.

\subsubsection{Statistical analyses.} 
We employed several statistical methods to analyze the survey data. To assess which type of discrepancy between individuals’ public and
private opinions was more commonly observed, we used the Wilcoxon signed-rank test with Holm correction for multiple comparisons as the data are paired and not normally distributed. For the analysis of topic-specific differences in opinion expression, we utilized the Spearman's rank correlation \cite{zar2005spearman}. We used logistic regression with standard errors clustered at the participant level to evaluate the probability of a participant choosing to remain silent, given that the dependent variable in this scenario is binary \cite{nick2007logistic}.
For analyses with an ordinal dependent variable, we employed ordered logistic regression with standard errors clustered at the participant level to examine the factors 
correlated with discrepancies between public and private opinions \cite{fullerton2009conceptual}. We then conducted mediation analyses for both models to investigate potential indirect associations \cite{mackinnon2007mediation}. 

\subsubsection{Thematic analyses.} 
Open-ended survey questions were coded using thematic analysis \cite{braun2006using}. 
Two authors served as the primary coders and were responsible for the initial coding and iterative refinement of the codebook. Both primary coders independently familiarized themselves with the data and conducted open coding on an initial subset of responses to identify salient concepts. Using the resulting codebook, they iteratively coded the full datasets, refined code definitions, and organized codes into higher-level thematic categories. Intercoder reliability was assessed using Cohen’s kappa, treating each category as a binary label. Because each item could be assigned multiple determinants, kappa was computed separately for each category. Kappa values ranged from 0.76 to 1.00 across categories, reflecting relatively high agreement. A third author acted in a review and adjudication role. After the coding and theme development were completed, the third author reviewed the codebook and thematic structure to assess consistency and coherence. Any remaining minor discrepancies were discussed among the authors and resolved through consensus.

\small
\begin{table*}
\centering
\scalebox{0.75}{
\begin{tabular}{cccccc}
\toprule
\textbf{Participant} & \textbf{Gender} &     \textbf{Age} &         \textbf{Race/Ethnicity} &   \textbf{Education} &    \textbf{Political Orientation} \\
\midrule
p1 & Transgender Female & 18 to 24 &        White/Caucasian & High school degree or equivalent&    Democrat \\
p2 &DemiMan & 25 to 34 & White/Caucasian & High school degree or equivalent & Independent \\
p3 & Male & 35 to 44 & White/Caucasian & High school degree or equivalent & Independent \\
p4 & Male & 45 to 54 & Asian & Associate degree & Independent \\
p5 & Female & 25 to 34 & White/Caucasian & Graduate degree&  Republican \\
p6 & Female & 35 to 44 & White/Caucasian &  Bachelor degree&    Democrat \\
p7 &Female & 45 to 54 & White/Caucasian & Bachelor degree& Independent \\
p8 &Female & 18 to 24 & Black/African American & High school degree or equivalent  & Independent \\
p9 &Female & 35 to 44 & White/Caucasian & Graduate degree &  Republican \\
p10 & Male & Above 55 &White/Caucasian & Bachelor degree&    Democrat \\
p11 & Male & 25 to 34 &Asian &  Bachelor degree & Democrat \\
p12 & Male & 25 to 34 & White/Caucasian &    Graduate degree &    Democrat \\
p13 & Female & Above 55 & White/Caucasian & Bachelor degree &    Democrat \\
p14 & Male & 45 to 54 & White/Caucasian & Bachelor degree & Independent \\
p15 & Male & 25 to 34 & White/Caucasian & High school degree or equivalent  &    Democrat \\
p16 & Female & 25 to 34 & White/Caucasian & Bachelor degree&    Democrat \\
p17 & Female & Above 55 & Black/African American &   Bachelor degree & Independent \\
p18 & Male & 35 to 44 & White/Caucasian & High school degree or equivalent  &  Republican \\
p19 & Male & 45 to 54 & White/Caucasian & Attending College&    Democrat \\
p20 & Female & Above 55 & White/Caucasian & Associate degree &    Democrat \\
\bottomrule
\end{tabular}
}
\caption{Interview participant demographics.}
\label{tab:interview_d}
\end{table*}
\normalsize

\subsection{Participant Interviews}

We conducted 20 semi-structured interviews with volunteers amongst the survey participants. All interviews were carried out via Zoom. Length of interviews ranged from 19 to 35 minutes; the majority were completed within 25 minutes. Interview questions are provided in Appendix 2. 

The survey results established systematic patterns in self-censorship and opinion discrepancy, describing how participants reported behaving across different community contexts. The interviews were conducted to complement these findings by examining how and why participants arrived at these decisions. In particular, the interviews allowed us to probe the reasoning processes underlying observed survey patterns, clarify how participants interpreted key survey constructs, and investigate tensions in the survey responses that could not be resolved through quantitative analysis alone. This mixed-method design enabled us to connect aggregate patterns to individual sense making around opinion expression.

\subsubsection{Interview design.} The semi-structured interviews were organized around five thematic areas. Each theme was selected to deepen understanding of specific patterns or open questions that emerged from the survey results.

\noindent \textbf{Reasons for discrepancies between public and private opinions.} Survey responses indicated that some participants reported substantial gaps between their privately held beliefs and their public expressions, while others reported little or no discrepancy. This interview section was designed to understand how participants interpreted these discrepancies, the situations in which they arose, and the reasoning participants used to justify expressing or withholding certain views. For participants who reported no discrepancy, questions focused on why they perceived alignment between private and public opinions and how they assessed the risks of expression.

\noindent \textbf{Determinants of opinion sharing.} Responses to the open-ended survey questions suggested that participants considered additional, more nuanced factors influencing opinion sharing that were not fully captured by the closed-ended measures. Interview questions in this section therefore explored how participants weighed these considerations in practice and surfaced other influences on expression decisions that were not fully captured in the survey.

\noindent \textbf{Crafting a public image.} Survey findings suggested that concerns about audience perception and self-presentation were linked to both silence and opinion adjustment. This interview section examined how participants thought about managing their public image on social media and how identity considerations influenced expression decisions.

\noindent \textbf{Influence of social media community.} While the survey captured community context (e.g., perceived audience size and support) at a high level, this section was designed to unpack how participants understood their community, how they inferred norms or expectations, and how community characteristics factored into their anticipation of feedback or judgment.

\noindent \textbf{Consequences and influence of social media dynamics.} Finally, the survey raised questions about reinforcement, conformity, and the broader consequences of opinion expression over time. This section invited participants to reflect on perceived positive and negative outcomes of opinion reinforcement and conformity, as well as how platform features and social dynamics shaped their longer-term expression strategies.

\subsubsection{Recruitment and compensation.} 
At the end of the survey, all participants were given the opportunity to sign up for a follow-up interview. The decision to invite survey participants to these interviews, rather than sourcing independent participants, was made because their responses to the survey's open-ended questions enabled us
to tailor our semi-structured interview questions to probe more deeply into the reasons behind their responses. 
A total of 42 survey respondents indicated interest in participating in a follow-up interview. From this pool, we used purposeful sampling \cite{patton2014qualitative,palinkas2015purposeful} to select participants who reflected diversity in opinion expression patterns, including individuals who reported consistent opinion sharing, selective silence, and some level of discrepancies between private beliefs and public expression. We aimed to maintain a balanced representation across these three categories (target ratio 1:1:1), and the final sample closely approximated this distribution (7:7:6). Selection also considered the clarity and substantive depth of participants’ open-ended survey responses to ensure that interviewees could articulate their reasoning and experiences in detail. For example, we excluded responses that were extremely short or lacked explanation (e.g., one-word or single-sentence answers without justification), as these provided limited insight into participants’ underlying reasoning. In contrast, participants who provided more elaborate responses with clear justification or examples were prioritized for inclusion.
Each interview participant was paid \$15 upon completion, which is higher than the minimum wage standard. In total, \$300 was used for interview participant compensation. Demographic information for interview participants is provided in Table~\ref{tab:interview_d}.

\subsubsection{Thematic analyses.} After conducting a quality check on the transcriptions automatically generated by Zoom, we analyzed interview transcripts using a thematic analysis approach via Nvivo-14 \cite{braun2006using,dhakal2022nvivo}. Two authors served as the primary coders, with a third author involved in reviewing and adjudicating the coding process. This collaborative and iterative approach is commonly used in Human-Computer Interaction research \cite{wu2024reacting,ammari2015understanding}. The analysis proceeded in multiple stages, including familiarization, open coding, iterative refinement of a shared codebook, and theme synthesis. Initially, the first author familiarized themselves with the data through in-depth reading of the transcriptions. During the open coding phase, initial codes were identified 
from participants' answers to the seed questions. These codes were grouped into potential themes relevant to the RQs. Following the initial coding, two authors collaboratively discussed the meaning, similarities, and differences among identified themes. They decided collectively to retain, remove, or reorganize these themes. 
These were then synthesized into broader themes that could represent patterns of opinion expression more comprehensively. Using this preliminary codebook, subsequent transcripts were coded iteratively. The research team held regular meetings to review coding decisions, clarify code definitions, and group related codes into higher-level categories. Disagreements were discussed among the coders until consensus was reached, and the codebook was refined throughout this process. After the primary coding and theme development were completed, a third author reviewed the existing codes, codebook, and thematic groupings to support the quality and coherence of the analysis. Any remaining minor discrepancies were discussed among the authors and resolved through consensus. The first and second authors served as the primary coders, while the third author acted in a review and adjudication role, focusing on code consistency and thematic coherence rather than independent recoding of all transcripts. All authors reviewed the resulting thematic structure and supporting excerpts. This iterative process resulted in a set of thematic clusters that form the basis of the qualitative findings reported in the paper.

\subsubsection{Effective sample size.}
Thematic saturation refers to the point in qualitative analysis at which additional data no longer yield substantively new themes \cite{hennink2022sample}. In this study, interviews and analysis proceeded iteratively, and a total of 20 interviews were conducted. At this point in the coding process, no substantively new themes emerged.

\subsection{Response Bias Control} 
To reduce the likelihood that Prolific participants could infer the study’s intent and adjust their responses accordingly, we implemented several safeguards during recruitment and data collection. The study description and materials were presented with neutral framing (e.g., the title and objectives used “Understanding the patterns of opinion expression on social media platforms”), which did not disclose the underlying hypotheses. In addition, interview questions were deliberately structured to be asked in both directions (for example, addressing both potential positive and negative consequences of different opinion expression patterns, as shown in the Appendix). This approach reduced the likelihood that participants would modify their responses to align with perceived expectations of the researchers.

\section{Results}
We present results in three parts, corresponding to our research questions. We first examine (R1) whether there is a discrepancy between privately held beliefs and publicly expressed opinions on social media. We then analyze (R2) how contextual factors, including topic, perceived community, and posting behavior, are associated with users’ willingness to share opinions across different topics. Finally, we report qualitative findings and design implications based on (R3) users’ perceived consequences of self-censorship and reinforcement.
\subsection{RQ1: Public vs. Private Opinions}
This subsection addresses RQ1, examining whether and to what extent individuals exhibit discrepancies between their privately held beliefs and publicly expressed opinions on social media.
As noted above, a series of survey questions were designed to measure: reported differences between users' public and private opinions on a variety of topics; users' tendencies toward conformity and self-censorship; and, users' tendencies toward reinforcement. 
These questions employed a 7-point Likert scale measuring participants' agreement (ranging from ``strongly disagree'' to ``strongly agree'') with a series of statements. The calculated mean of the responses within each set yielded an agreement score for each participant, with potential values extending from -3 to 3. 

\begin{figure*}[ht]
    \centering
    \includegraphics[width=\linewidth]{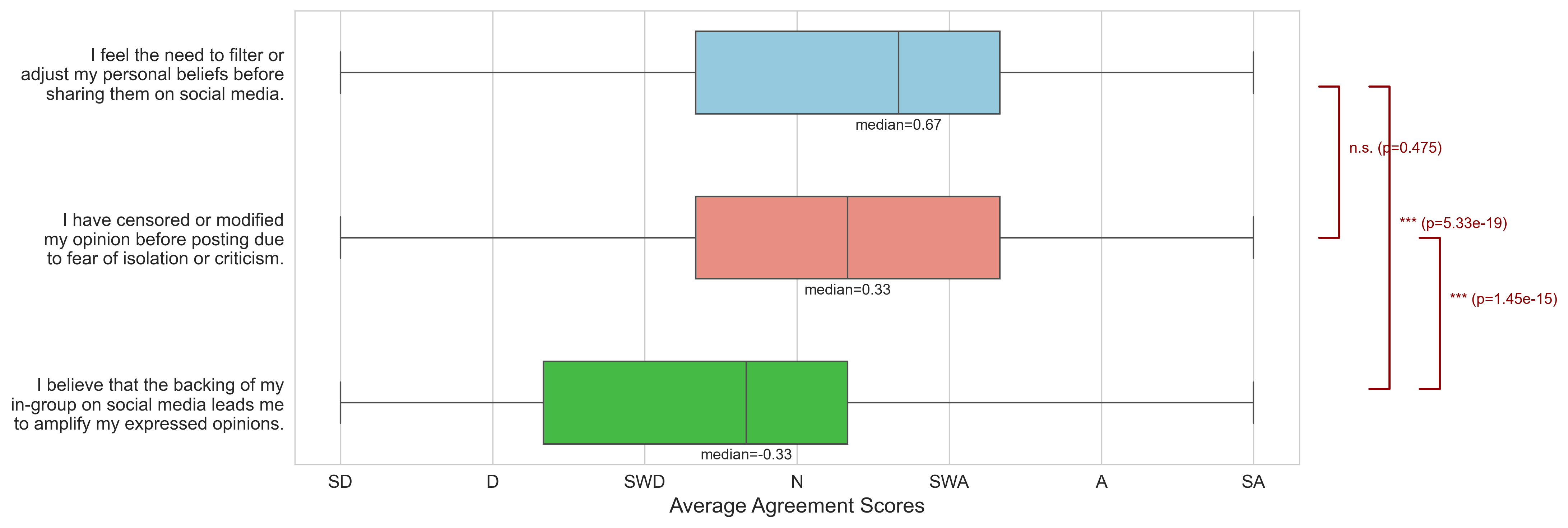}
    \caption{Participants' agreement with statements intended to capture: \emph{discrepancies} between private and public opinion; the extent to which participants experience \emph{conformity} or self-censorship; and, \emph{reinforcement} of their opinions within public social media discourse. An exemplar statement is provided for each of the three categories. The reported statistics are an average of three similar statements within the survey (see \emph{Opinion Expression} in Appendix 1: Survey Instrument). (SD: Strongly Disagree, D: Disagree, SWD: Somewhat Disagree, N: Neither Agree Nor Disagree, SWA: Somewhat Agree, A: Agree, SA: Strongly Agree, p-value: p-value from Wilcoxon test.)}
    \label{fig:discrepancy}
\end{figure*}

Response distributions and their corresponding Wilcoxon test p-values are shown in Figure~\ref{fig:discrepancy}. Participants confirmed that there exists a discrepancy between their privately held beliefs and publicly expressed opinions. They also agreed, on average, that they feel an inclination to conform when expressing their public opinions. 
A minority of participants (29.23\%) report amplifying their true opinions on social media platforms. 

\subsection{RQ2: Contextual Influences on Opinion Expression}
To address RQ2, we systematically explore context along multiple axes. Our analyses are structured into three parts. The first explores relationships between topics of conversation and participants' willingness to share their opinions on social media. Second, we model the effects of various individual and community-level features on opinion expression, including users' perceived community support and community size. Finally, we extract additional contextual factors raised by participants during open-ended survey questions and interviews to construct a holistic understanding of influences on opinion expression.

\begin{table}[ht]
\centering
\scalebox{1}{
\begin{tabular}{cccc}
\hline
\textbf{Topic} & \textbf{\(\rho\)} & \textbf{P-value} & \textbf{95\% CI} \\ \hline
Politics & 0.16 &$< 0.01^{**}$   &[0.06, 0.25] \\
 Social Justice&0.16&$ < 0.01^{**}$& [0.06, 0.25]   \\
     Health &0.18&$ < 0.001^{***}$&[0.08, 0.28]   \\
     Environment&0.28&$ < 0.001^{***}$&[0.19, 0.37]   \\
        Technology&0.25&$ < 0.001^{***}$&[0.16, 0.34]   \\
        Religion &0.31&$ < 0.001^{***}$&[0.22, 0.40]   \\
\hline
\end{tabular}
}
\caption{Topic-wise Spearman correlations between polarization and opinion expression, with p-values and confidence intervals.}
\label{table:polarity_willingness}
\end{table}

\begin{figure*}[ht]
    \centering
    \includegraphics[width=0.9\linewidth]{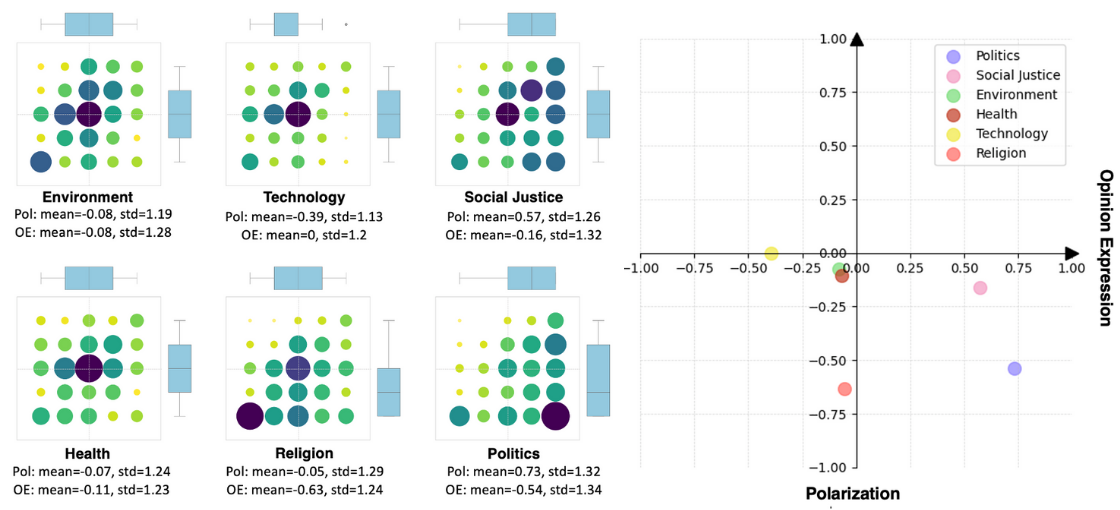}
    \caption{Perceived polarization and willingness to express true opinion, by topic (Pol=Polarization, OE=Opinion Expression).}
    \label{fig:polarityvsexpression}
\end{figure*}
\subsubsection{Categories and polarities of topics.}
\label{polvexp}
We asked survey participants to rate the polarity of our six focal topics on a 5-point Likert scale. Likewise, we asked them to rate their willingness to post their true opinions about each topic on social media. 
As the data is discrete and ordered, we performed Spearman rank-order correlation to quantify the relationship between topic polarity and opinion expression (see Table \ref{table:polarity_willingness}). We observe that for less polarizing topics, such as technology, environment, health, and religion, the correlation between opinion expression and perceived polarization is significantly positive. That is, people are more likely to engage in discussions that are perceived as more controversial. This effect is diluted in discussions of politics and social justice as many participants are less willing to express their true opinions due to the high level of perceived topic polarity (see Figure~\ref{fig:polarityvsexpression}). Specifically, we observe a greater concentration of high polarization and low opinion expression scores, particularly evident in the bottom right corner, for topics related to social justice and politics. When plotting average polarization vs. opinion expression across all six topics (see Figure~\ref{fig:polarityvsexpression}, right), a consistent pattern emerges: topics perceived as more polarizing by the public demonstrate a lower overall opinion expression and higher levels of conformity. The contrast between the within-topic and between-topic patterns suggests that the relationship between perceived polarization and willingness of opinion expression is non-monotonic and context-dependent: within less sensitive topics, perceived polarization may stimulate engagement; however, for topics that are already highly divisive, a substantial portion of individuals choose to suppress expression, potentially due to heightened conformity pressures. 

Thematic analysis of the open-ended survey questions (see Table~\ref{tab: survey_topic} in the Appendix) and interviews supported these findings. Responses to the open-ended questions frequently referenced topics and subtopics that aligned with the six topical categories used in our analysis, indicating that these categories capture a broad range of situations in which self-censorship and opinion conformity arise. Several participants explicitly stated their reluctance to engage with polarizing topics on social media.

\begin{quote}
    \textit{I have been really hesitant to post at all about Palestine because it is so incredibly polarizing and people seem angry on both sides if you post. I am afraid of being called antisemitic for supporting Palestine. I have been posting more mild versions of my opinion instead of calling for a ceasefire. }--Survey participant
\end{quote}

\subsubsection{Variable operationalization.}

To examine how community context shapes self-censorship, we constructed two regression models. \textbf{Model 1} examines the decision to remain silent (binary outcome), 
while \textbf{Model 2} examines opinion discrepancy among those who chose to post (ordinal 
outcome). Both models use the same independent variables but capture different 
aspects of self-censorship. The two dependent variables correspond to distinct stages of opinion expression: the decision to speak and the degree of adjustment in expressed opinions. Table \ref{tab:variables} summarizes all variables and their operationalization\footnote{Because our data are observational and cross-sectional, all estimated relationships, including those involving independent variables, mediators, and interaction terms, are interpreted as associational patterns rather than as causal effects. References to direct or indirect effects reflect statistical decomposition of associations rather than causal transmission.}. The selection of the target variables is theoretically grounded and introduced in Section 2.1.
\begin{table}[H]
\tiny
\centering
\begin{tabular}{p{1.2cm}p{2.5cm}p{9.2cm}}
\hline
Variable Type & Variable & Operationalization\\
\hline
Dependent & Decision to remain silent & 
Binary indicator of whether a participant reports willingness to share an opinion on a given topic (1 = silent, 0 = expresses opinion). This outcome is constructed from survey responses to topic-specific questions asking about intention to post. \\

Dependent & Opinion discrepancy & 
Ordinal measure of the difference between privately held beliefs and publicly expressed opinions among participants who report willingness to post. Both are measured on a 7-point Likert scale and recoded to the range [-3, 3]. Discrepancy is computed as $|\text{True Opinion}| - |\text{Public Expression}|$. \\

Independent & Perceived polarity & Participant-rated topic controversy (5-point scale). Section~\ref{polvexp} further documents topic-dependent patterns of expression in this study. \\
Independent & Perceived support & Percentage of connections perceived to share the participant’s view.  \\
Independent & Community size & Self-reported total number of connections on primary platform. \\
Mediator & Posting frequency & Self-reported posting frequency. Empirically, posting frequency shows a moderate correlation with community size (0.35) (Tables~\ref{tab:cm_1} and \ref{tab:cm_2}), indicating that posting frequency is meaningfully related to audience scale. \\
Control & Topic & Six predefined topic categories are included to adjust for topic-specific differences in sensitivity and baseline expression patterns that may confound associations between contextual variables and outcomes. \\
Control & Political ideology & Democrat / Republican / Neutral. Political ideology is included to adjust for observed imbalance in participant distribution and to account for baseline differences in opinion expression associated with partisan orientation. \\
Interaction & Topic $\times$ Polarity & Interaction term capturing topic-dependent effects. This interaction captures topic-dependent variation in the association between perceived polarity and opinion expression. As shown in Section~\ref{polvexp}, higher polarity is associated with increased engagement for some topics and reduced expression for others.  \\
\hline
\end{tabular}
\caption{Model variables, operationalization, and theoretical rationale. The correlation matrices (see Table~\ref{tab:cm_1} and \ref{tab:cm_2} in Appendix 3) show that all three independent variables have correlation coefficients lower than 0.1, suggesting that the three dimensions capture non-redundant information. Variance Inflation Factor (VIF) values were low (shown in Table~\ref{tab:vif_models} in Appendix 3, all $<$5), indicating no significant multicollinearity among the independent variables.}
\label{tab:variables}
\end{table}

\subsubsection{Model 1: Impacts of community factors on self-censorship.}

\begin{table}[ht]
\small
\centering
\begin{tabular}{m{3.5cm}m{1.2cm}m{1.2cm}m{1.1cm}m{1.5cm}m{1.6cm}m{0.9cm}}
\hline
\textbf{Variable} & \textbf{Coef.} & \textbf{Std. Err.} & \textbf{z} & \textbf{P$>|z|$} & \textbf{95\% CI} & \textbf{Odds Ratio} \\ 
\hline

Support & -0.39 & 0.08 & -5.19 & $< 0.001^{***}$ & [-0.53, -0.24] & 0.68 \\

Polarization & -0.16 & 0.21 & -0.78 & 0.43 & [-0.57,  0.24] & 0.85 \\

Posting Frequency & -0.54 & 0.23 & -2.31 & 0.02* & [-0.99, -0.08] & 0.58 \\

Community Size &  0.19 & 0.09 &  2.04 & 0.04* & [ 0.01,  0.37] & 1.21 \\

Topic: Health &  0.18 & 0.20 &  0.87 & 0.39 & [-0.22,  0.57] & 1.19 \\

Topic: Politics &  0.00 & 0.25 &  0.01 & 0.99 & [-0.49,  0.49] & 1.00 \\

Topic: Religion &  0.10 & 0.19 &  0.54 & 0.59 & [-0.27,  0.48] & 1.11 \\

Topic: Social Justice &  0.11 & 0.20 &  0.54 & 0.59 & [-0.28,  0.49] & 1.11 \\

Topic: Technology &  0.06 & 0.17 &  0.34 & 0.73 & [-0.27,  0.38] & 1.06 \\

Democrat (vs. Neutral) &  0.37 & 0.35 &  1.05 & 0.29 & [-0.32,  1.06] & 1.45 \\

Republican (vs. Neutral) &  0.35 & 0.35 &  0.99 & 0.32 & [-0.34,  1.04] & 1.42 \\

Polarization $\times$ Health &  0.24 & 0.24 &  1.00 & 0.32 & [-0.23,  0.71] & 1.27 \\

Polarization $\times$ Politics &  0.56 & 0.30 &  1.88 & 0.06 & [-0.02,  1.15] & 1.76 \\

Polarization $\times$ Religion &  0.19 & 0.23 &  0.85 & 0.40 & [-0.25,  0.63] & 1.21 \\

Polarization $\times$ Social Justice &  0.16 & 0.24 &  0.66 & 0.51 & [-0.31,  0.62] & 1.17 \\

Polarization $\times$ Technology &  0.09 & 0.16 &  0.57 & 0.57 & [-0.22,  0.40] & 1.09 \\

\hline
\end{tabular}
\caption{Logistic regression results predicting self-censorship (Model~1; $n=2170$). Odds ratios are reported to indicate effect sizes, alongside coefficients, clustered standard errors, and 95\% confidence intervals.} 
\label{tab:logit_results}
\end{table}

We employ logistic regression on 2,170 out of 2,340 (6x390) valid survey responses to questions asking participants about their intention to share an opinion on a variety of topics. The dependent variable is defined as a binary indicator of whether a participant remains silent or expresses an opinion, as summarized in Table~\ref{tab:variables}.
Our analysis reveals that both perceived community support and posting frequency are negatively correlated with the decision to remain silent (Table~\ref{tab:logit_results}). This suggests that social media users are more inclined to remain silent if they perceive a lack of support within their community or if they are infrequent users. Conversely, community size is positively associated with silence, indicating that users who believe they are exposed to a larger audience are more likely to refrain from expressing their opinion, possibly due to increased concerns about social judgment or backlash, consistent with self-silencing theory.
We also observe a marginally significant interaction between perceived polarization and the topic of politics, indicating that polarization may have a stronger effect on self-censorship when the topic is political. This suggests that the relationship between perceived polarization and willingness to express opinions could potentially be context-dependent and intensified in political discussions. 

To further examine this relationship, we conducted a mediation analysis to test the indirect pathway from community size to posting frequency, and subsequently to the decision to remain silent (Table~\ref{tab:mediation_1} in Appendix 3). We found that larger community size is positively associated with more frequent posting (path coefficient = 0.33, 95\% CI = [0.29, 0.45]), and that posting frequency significantly reduces the likelihood of remaining silent (path coefficient = –0.53, 95\% CI = [–0.98, –0.07]). This results in a significant negative indirect effect (–0.18, 95\% CI = [–0.22, –0.04]), indicating that increased posting partially offsets the silencing effect of larger audiences. At the same time, the direct effect of community size on self-censorship remains positive and significant (0.19, 95\% CI = [0.03, 0.38]), consistent with the main model. These findings indicate that community size is associated with self-censorship both directly and through changes in posting frequency.

\subsubsection{Model 2: Discrepancies amongst expressed opinions.}

Our second analysis examines associations between the set of factors listed in Table~\ref{tab:variables} and differences between privately held true beliefs and publicly expressed opinions, focusing on expression outcomes among participants who report willingness to share. The operationalization of this dependent variable is summarized in Table~\ref{tab:variables}. Disparity y is calculated as: 
$ \text{Disparity}(y) = \left| \text{True Opinion} \right| - \left| \text{Public Expression} \right|$, where both quantities are measured on a 7-point Likert scale and recoded to the range [-3, 3]. The use of absolute values ensures that any reduction in the strength of opinion—regardless of positive or negative view—is treated as positive disparity. That is, whether a participant moderates a strong view (e.g., True:-3; Public: -1) or reinforces a weak view (e.g., True:-1; Public: -3), the disparity reflects the magnitude and direction of the change. Importantly, our design intentionally restricts directional reversals between private and public opinions (cases where the private and public opinions are completely opposed) to ensure that disparity reflects changes in intensity, not strategic misrepresentation or deliberate opinion reversal, as these fall outside the conceptual scope of opinion conformity and reinforcement.

We have 1,627 out of 2,340 (6x390) valid responses from participants reporting willingness to share their opinion on a particular topic. We employed ordinal logistic regression with standard errors clustered at the participant level and the result is shown in Table~\ref{tab:ordered_model_results}. We note that a larger absolute value in the discrepancy signifies a greater divergence between private belief and public expression. 

\begin{table}[ht]
\small
\centering
\begin{tabular}{m{3.5cm}m{1.2cm}m{1.2cm}m{1.1cm}m{1.5cm}m{1.6cm}m{0.9cm}}
\hline
\textbf{Variable} & \textbf{Coef.} & \textbf{Std. Err.} & \textbf{z} & \textbf{P$>|z|$} & \textbf{95\% CI} & \textbf{Odds Ratio} \\ 
\hline

Support & -0.13 & 0.06 & -2.06 & 0.04* & [-0.25, -0.01] & 0.88 \\

Polarization & 0.01\ & 0.13 & 0.08 & 0.93 & [-0.24, 0.26] & 1.01 \\

Posting Frequency & -0.02 & 0.06 & -0.37 & 0.71 & [-0.15, 0.10] & 0.98 \\

Community Size & 0.09 & 0.05 & 1.70 & 0.09 & [-0.01, 0.19] & 1.09 \\

Topic: Health & 0.04 & 0.18 & 0.24 & 0.81 & [-0.30, 0.39] & 1.04 \\

Topic: Politics & 0.03 & 0.24 & 0.10 & 0.92 & [-0.44, 0.50] & 1.03 \\

Topic: Religion & -0.05 & 0.17 & -0.27 & 0.79 & [-0.39, 0.29] & 0.96 \\

Topic: Social Justice & -0.07 & 0.22 & -0.33 & 0.74 & [-0.49, 0.35] & 0.93 \\

Topic: Technology & -0.06 & 0.15 & -0.41 & 0.68 & [-0.36, 0.24] & 0.94 \\

Democrat (vs. Neutral) & -0.01 & 0.20 & -0.07 & 0.94 & [-0.41, 0.38] & 0.99 \\

Republican (vs. Neutral) & 0.05 & 0.20 & 0.26 & 0.80 & [-0.34, 0.45] & 1.05 \\

Polarization $\times$ Health & 0.07 & 0.18 & 0.36 & 0.72 & [-0.29, 0.43] & 1.07 \\

Polarization $\times$ Politics & 0.10 & 0.26 & 0.39 & 0.70 & [-0.40, 0.60] & 1.10 \\

Polarization $\times$ Religion & 0.01 & 0.19 & 0.04 & 0.97 & [-0.37, 0.38] & 1.01 \\

Polarization $\times$ Social Justice & -0.02 & 0.23 & -0.10 & 0.92 & [-0.47, 0.43] & 0.98 \\

Polarization $\times$ Technology & -0.05 & 0.15 & -0.32 & 0.75 & [-0.34, 0.24] & 0.95 \\

\hline
\end{tabular}
\caption{Ordered logistic regression results predicting the degree of opinion expression discrepancy (Model~2; $n=1627$). Odds ratios are reported to indicate effect sizes, alongside coefficients, clustered standard errors, and 95\% confidence intervals.}
\label{tab:ordered_model_results}
\end{table}

We find a negative association between perceived support and the predicted probability of opinion discrepancy, with lower perceived support corresponding to greater discrepancy.
Mediation analysis (Table~\ref{tab:mediation_2}, Appendix 3) provides further insight: while the direct effect of community size is not statistically significant, the indirect effect through posting frequency is negative and significant (–0.03, 95\% CI = [–0.08, -0.00]). This indicate a weak indirect relationship in which posting frequency links community size to differences between private beliefs and public expression.

Taken together, findings from Model 1 and Model 2 address two related aspects of opinion expression. First, social media users in larger online communities, with lower perceived support and lower posting frequency, are more likely to remain silent. Mediation 
analysis shows this effect operates both directly and through posting frequency. 
Second, among users who do express opinions, lower perceived support is associated 
with greater discrepancies between private beliefs and public expression, suggesting 
adjustment to align with perceived group norms or maintain social cohesion.

\subsubsection{Additional determinants of opinion expression.}
While our models captured structural predictors that are quantifiable, participants also described additional factors influencing their willingness to express opinions—many of which may function as latent confounders. 
We further qualitatively identify and discuss key determinants of opinion expression from open-ended survey questions (see Table~\ref{tab:survey_determinants}) and interview responses.

\begin{table*}
\centering
\begin{tabular}{p{3cm}p{6.9cm}p{0.9cm}p{1.5cm}}
\toprule
\textbf{Category} & \textbf{Subcategory} &  \textbf{Count}& \textbf{Percentage}\\
\midrule
Topic and Polarity& --- & 27&6.9\%  \\
Self-presentation& Desire for acceptance (59); Avoiding Conflict (27); Career Constraint (9); Credibility (5) &  100 &25.6\% \\
Evidence Support&  ---& 19&4.9\%  \\
Platform Affordance& --- &  4 &1.0\%\\

Social Influence& --- &  15&3.8\% \\
None&No Disparity or Posts (171); No Reason Given (64)&235&60.3\%\\
\bottomrule
\end{tabular}
\caption{Determinants of opinion expression for survey respondents (from open-ended questions). Note: Total percentage exceeds 100\% as responses could reference multiple determinants.}
\label{tab:survey_determinants}
\end{table*}

\noindent\textbf{Self-presentation.}
Social media users often engage in intentional and unintentional curation of their online personas through their choices about what to post and what not to post. This curation is driven by an awareness of how their self-image is perceived on social media, which acts as a barrier to full disclosure of their real identities. Users are conscious of the balance between authenticity and the need to maintain a certain image that aligns with their self-perception or the expectations of their communities. 

\begin{quote}
\textit{I was worried about how people who think they know me will perceive me based on certain opinions.} --P14

\textit{I just don't want to be coming off as somebody that shoots out their mouth without thinking about how it's gonna affect other people.} --P20
\end{quote}

Self-presentation is influenced by many factors including: fear of judgment or isolation (11 out of 20 participants); desire for acceptance (P06, P07); inherent cultural identities (P11,P20); and potential social repercussions that might follow from sharing true opinions (P04, P10, P12). 
Furthermore, the decision by some social media users to amplify or authentically express their opinions, regardless of external perceptions, can stem from a conscious awareness of their self-image. For instance, these individuals may seek to present themselves as: challengers of superiority complexes (P18); proponents of authenticity (P17); supporters of minority groups (P02, P14); and as advocates intent on making their voices heard (P02). 

We also observe that maintaining a coherent self-image can drive seemingly opposite actions. In some instances, the act of expressing extreme views can be a form of conformity to community norms. For instance, one participant shared that he posted strong stances on the Black Lives Matter movement because his social circle expected it of him. To maintain a positive image within the community, he felt compelled to present himself as a vocal supporter, despite not holding strong personal opinions on the issue. Reflecting conformity through extremism arises from the fundamental desire for social acceptance and identity management. 

\noindent\textbf{Evidence support.} The accuracy and truthfulness of information significantly influence participants' willingness to share opinions. For example,
\begin{quote}
     \textit{If I say something has happened, I make sure it's factual. I don't like propaganda, lies or misinformation.} --P13
\end{quote}
This reflects a broader concern about the integrity of information shared online and a personal commitment to contribute positively by ensuring the authenticity of shared content. These attitudes appear to be a deliberate effort to combat the spread of false information, emphasizing a possibly proactive approach to fostering a healthier environment.

\noindent\textbf{Platform norms and affordances.}
The nature of a platform fundamentally shapes users' willingness of opinion expression. We observe that social media users' willingness to express opinions is platform-dependent due to differences in audience composition, community norms, and platform-specific cultures. 
\begin{quote}
\textit{I'm not reluctant to express an opinion in general, just on social media. The platform is the key variable.} --Survey participant
\end{quote}
For instance, platforms like LinkedIn were described as spaces where expressing personal beliefs could risk professional consequences:

\begin{quote}
\textit{I use mainly Linkedin and I would not want anyone to see my personal beliefs or thoughts because that could have a negative impact on my ability to be hired or how future employers or current coworkers view me.} --Survey participant
\end{quote}

In contrast, platforms that are built upon weak identity ties, such as Reddit, were described as more flexible spaces for expression.  

\begin{quote}
\textit{If I were more anonymous I would probably engage more, at least on Reddit.} --P01
\end{quote}

Restrictions and moderation on social media platforms also shape user expression. Concerns about being ``punished'' for saying certain words reflect growing user awareness and dissatisfaction with the perceived overreach of platform moderation policies.
\begin{quote}
\textit{I noticed it on YouTube. People have shown that they can't say certain words, or they get punished for it. And to me, it seems like we're going towards that type of society where you can't say what you feel.} --P17
 \end{quote}

Affordances for anonymity can also encourage users to express themselves more freely, particularly about sensitive topics. However, they also raise questions about accountability in civil discourse, as users may engage in behavior they would avoid using their real identities.  
\begin{quote}
\textit{That's why people like Reddit so much, you post anonymously...There are things that you wouldn't really be comfortable sharing, because, you have that level of anonymity on there.} --P15
\end{quote}

\noindent\textbf{Social influence.}
Participants report sharing their opinions when they believe it might make a positive difference in some way. For example, the fear of misinformation spreading unchecked on social media platforms motivates some users to speak out. 
\begin{quote}
\textit{If I didn't say that then I would be afraid that other people who come across this comment might think that it's true. Especially on the Internet, we're so used to seeing false information. So even if I had a polarizing point of view, I would say it simply for educational purposes. }--P08

\textit{I don't exaggerate my ability to influence the world. But it's like the famous Indian saying, ripple in the pond starts with one rock in the middle,  and it expands out so you don't know what that ripple might do.} --P13
\end{quote}
Conversely, some participants believe that not engaging is the best strategy when comments or discussions are not likely to be constructive or change the situation. 
\begin{quote}
\textit{Engaging in said conversation or behavior wouldn't change anything at all, and it could just keep the ball rolling. The easiest way to get it to stop is not even acknowledging it.} --P18
\end{quote}

\noindent\textbf{Cultural and societal identity.}
This connection between personal identity and online behavior underscores the deeply personal nature of social media interactions, where users select and prioritize issues that reflect and affect their own lives and those of their communities.
\begin{quote}
\textit{I think the more personal to me will make me more willing to share it. So something that affects my identity and the identity of the people that I follow. So religion is a very personal one, that'll affect me nationality and my ethnic origin.} --P11
 \end{quote}

\subsection{RQ3: Consequences and Possible Interventions}
Building on our examination of the determinants of opinion sharing, this section explores users' perceptions of social media and its wide range of negative consequences. We discuss potential actions to improve the design of social media systems, mitigate these negative effects, and promote healthier digital environments.


\subsubsection{Consequences of conformity.}

Although the majority of participants reported experiencing conformity when sharing opinions on social media, 
they also noted the inherent challenge of observing this behavior in others.
\begin{quote}
    \textit{I see lots of extremism. If someone is moderating their opinion, I wouldn't really know it, because they're posting something that I'll take at face value.} --P09
\end{quote}
Nonetheless, participants acknowledged the negative effects of conformity. Through thematic analysis, we categorized users' reported consequences into user-level, community-level, and society-level impacts. The corresponding summary of the consequences from survey open-ended questions is shown in Table~\ref{tab:oq_consequences}.

\small
\begin{table}
\centering

\begin{tabular}{m{6cm}m{1cm}m{1.5cm}}
\toprule
\textbf{Category} &  \textbf{Count} &\textbf{Percentage}\\
\midrule
No Negative Consequences& 111 &28.5\% \\
Personal Discomfort&  98&25.1\%\\
 Suppression of Rational Discourse& 88&22.6\% \\
Unintended Encouragement of Extremism& 29&7.4\% \\
Blank or N/A & 44 &11.3\%\\
Depends/Maybe & 26&6.7\%\\
\bottomrule
\end{tabular}
\caption{Consequences of disparity for survey respondents (from open-ended questions). Note: Total percentage exceeds 100\% as responses could reference multiple consequences.}
\label{tab:oq_consequences}
\end{table}
\normalsize

\noindent \textbf{Personal discomfort.}
One participant highlighted an often overlooked yet significant effect of conformity on individual social media users. This effect is described by the psychological theory of cognitive dissonance, which refers to the discomfort that arises when one's actions conflict with their beliefs \cite{bai2019exploring}.
\begin{quote}
\textit{...there might be some cognitive dissonance from pretending you're going along with something if you don't. Again, this is online. It's not like if someone knocked on your door and said, `Here's a gun. What do you think?' where you might speak the truth.} --P09
\end{quote}

\noindent \textbf{Suppression of rational discourse.}
Conformity in social media discussions can lead to the dominance of one viewpoint, effectively silencing alternative perspectives. This suppression reduces opportunities for back-and-forth discussions that are essential for generating reasoned arguments and fostering rational debate on various topics. As a result, the overall quality of discourse diminishes.
\begin{quote}


\textit{The internet will have less reasonable people, because more reasonable people will feel like 'I can't express myself', and so they'll slowly stop posting.} --P08


\textit{It's one less voice towards a cause that matters. That voice might have an impact or might not have an impact, but you won't know. And at the end of the day you will have to live with the fact that you didn't speak out for others when it was needed.} --P11
\end{quote}

\noindent \textbf{Unintended encouragement of extremism.}
Many participants expressed concern that conformity may unintentionally foster extremism by not actively opposing harmful content. This passive acceptance creates an environment where extreme views can flourish unchecked and silence can be interpreted as endorsement. 
\begin{quote}
\textit{Silence is violence. If you're not speaking up against something, you're just as complicit as the person committing the act.} --P02


\textit{You're not voicing your side, and it could cause them to believe whatever they want, or even possibly causes them to do something more extreme to get more attention.} --P18
\end{quote}


\subsubsection{Actions to reduce conformity.}
Participants identified several actions for content promotion and moderation systems on social media platforms, aiming to address conformity and self-censorship and support a more open and constructive online environment. These strategies aim to mitigate the influence of pressures towards conformity and promote healthier engagement among users.

\noindent \textbf{Fact-checking to support evidence-based expression.}
As observed, individuals prone to conformity are more likely to post when they have data or news to support their claims. While many social media platforms have introduced fact-checking features—such as labeling disputed content, linking to credible sources, or flagging misinformation—these practices have shown mixed results \cite{clayton2020real,walter2020fact}. Nonetheless, participants highlighted the potential of well-implemented fact-checking systems to reduce conflict and boost users' confidence in posting truthful content.
 
\begin{quote}
    
\textit{At some point they started doing fact checking when people were posting stuff, and there would be alerts like `no, you can't post that because it's not factual'... And if I can post something in it and it's proven to be factually based, I'm probably a lot more likely to post it. --P16}
\end{quote}

These reflections suggest that enhancing the visibility, transparency, and responsiveness of built-in or external fact-checking mechanisms could empower users to participate more actively and reduce hesitation driven by fear of spreading misinformation or being challenged unfairly.

\noindent \textbf{Providing incentives for healthy interactions.}
Efforts to design moderation systems often focus on demoting extremism \cite{ganesh2020countering,clifford2021moderating,lakomy2023online}, but less attention is given to incentivizing users who engage in reasonable and productive conversations. Participants highlighted the importance of addressing the conformity effect by implementing targeted incentives.

\begin{quote}
\textit{I think they could offer more incentives, and they should have more specified rules and take serious actions against people who do these type of things instead of promoting discussion boards where people are attacking others and making it go viral... I think that would do good for social media and make people more comfortable.} --P08
\end{quote}

\noindent \textbf{Fair interventions to counteract extremism.}
Many participants indicated that a reason for their self-censorship is the highly polarized environment and the need to protect themselves from extremism. This situation is exacerbated by the lack of consequences for extreme behaviors online, and such extremism often leads to greater monetization \cite{ballard2022conspiracy}.

\begin{quote}
\textit{They're refraining from talking about sensitive subjects because of the extreme thoughts. They're in the vileness of how people handle online interactions because there are no consequences online. People can say whatever, they know they can walk away clean from it.} --P15

\textit{It draws attention...They get a lot of interaction in their post, and if they're monetizing, the more clicks they get, the more money they can make.} --P13
\end{quote}

\noindent Addressing this issue by counteracting the reinforcement of extreme opinions could, in turn, help alleviate the need for self-censorship (P09). 

\begin{quote}
\textit{I think accountability prevents hate speech. You can also have more of a conversation.} --P11
\end{quote}

Moreover, the problem extends beyond platform self-regulation, as some platforms inherently encourage extreme content due to monetization incentives. Participants explicitly pointed out the need for external regulatory oversight:

\begin{quote}
\textit{I think there should be more regulation of big tech in general. Because the reality of it is that I think it's done a lot of things to our society that are not good, and quite honestly begging them to be nice about it hasn't really worked so at some point somebody else has to step in and do it.} --P10
\end{quote}

These insights underscore that addressing online extremism effectively requires a balanced approach combining robust internal moderation mechanisms with external regulatory frameworks to ensure accountability and reduce incentives for polarizing content.

\noindent\textbf{Promoting free speech with platform–community collaboration.}
To further encourage user participation and mitigate the spiral of silence, platforms should reconsider their policies and mission statements to explicitly advocate for open dialogue and inclusivity. Clearly articulated commitments such as declarations emphasizing the welcome of diverse viewpoints can reassure users about their freedom to communicate openly. Additionally, refining moderation practices by providing greater transparency around user bans or topic restrictions could build trust and encourage users to express their opinions more freely.
\begin{quote}
\textit{We allow everyone to talk, but maybe then loosening how they might ban people or ban topics like on Reddit, or just being more upfront with saying, hey, we're inclusive. Go ahead and talk about whatever you want to talk about here. So then people can open up.} --P12
\end{quote}

However, promoting free speech effectively involves more than modifying platform-level policies; community dynamics also play a significant role. Another participant emphasized:

\begin{quote}

\textit{I think it's more about the community than the platform. And I don't think everyone has to have an opinion on every issue. But I feel like some issues are more pressing than others and maybe if there was a way for a platform to make it easier for people to express their opinions, then that'd be cool.} --P11

\end{quote}

This perspective underscores the necessity for platforms and communities to work together to foster user engagement. Platforms could support this collaboration through targeted design features, such as dedicated spaces or prompts that highlight critical societal topics.

\subsubsection{Consequences of reinforcement.}
While conformity pressures limit users' ability to express genuine opinions openly, reinforcement further intensifies these dynamics by amplifying homogeneous perspectives and silencing opposite voices. Participants highlighted critical consequences that arise from these reinforcement effects, including echo chambers and incitement to harm, that indirectly promote conformity and intensify the negative impacts previously identified.

\noindent\textbf{Echo chambers and polarized groups.}
The most frequently mentioned consequence of reinforcement is the creation and maintenance of echo chambers. Such echo chambers isolate individuals within homogeneous communities, significantly narrowing their exposure to diverse perspectives and reinforcing polarized and often biased beliefs \cite{cinelli2021echo}. Participants noted that this not only diminishes critical thinking but also amplifies societal division.
\begin{quote}

\textit{The bad thing is that people will get a false sense of reality. So if your entire school community, teachers, and family—if everybody thinks the same way, you have no connection to reality because you're living in an echo chamber. ...  It is bad for objective reality because it is harder to understand the points of view of people who disagree with you.} --P09
\end{quote}

\noindent\textbf{Incitement to harm and violence.}
Reinforcement of extreme views can also lead to increased potential for real-world harm, including misinformation and aggressive behaviors. Participants stressed that continuous affirmation of extremist or harmful opinions on social media can normalize aggressive attitudes, which might manifest in offline violence or harmful actions.

\begin{quote}
\textit{Because when you're hiding behind a screen, you don't know how something you said is going to affect somebody else. You don't know what's going on in their life. What if their comment is the thing that drove them over the edge, whether it's suicide, or doing something crazy like shooting up a school.} --P06
\end{quote}

Such harmful behaviors and attitudes reinforce conformity indirectly, by creating a hostile environment where users fear backlash or aggressive responses if they disagree or challenge the majority view, further limiting open expression and increasing users' likelihood of self-censoring.
\begin{quote}
\textit{One of the rules I kind of made after those experiences was to refrain from commenting on people that are less close to me unless I absolutely know them well.
} --P10
\end{quote}

\subsubsection{Actions to avoid reinforcement.}
Participants repeatedly acknowledged a significant interplay between reinforcement and conformity: the stronger the reinforcement of dominant views, the stronger the pressure to conform and the less space for rational, diverse discussions. In response to the intertwined challenges of conformity and reinforcement, participants recommended several actions to mitigate reinforcement effects that extend upon the efforts previously described for reducing conformity. 

\noindent\textbf{Exposure to diverse opinions.}
A commonly suggested approach involves algorithmically ensuring exposure to diverse viewpoints to help users step outside their echo chambers and engage with broader perspectives.
\begin{quote}
\textit{Maybe some sort of algorithmic change so that they are still able to see things that are not just in their circle.} --P12
\end{quote}
Current recommendation systems inadvertently reinforce echo chambers by primarily presenting content users already prefer, which is driven by monetization and engagement rather than encouraging balanced discourse. To address this, recommendation algorithms should intentionally introduce verified content representing alternative or opposing viewpoints. Such design would actively disrupt echo chambers and provide users with broader perspectives, helping counteract both reinforcement and conformity pressures.

\noindent\textbf{Content creator training and education.}
Participants also suggested platforms could offer training or education programs to enhance content creators' abilities to produce responsible and meaningful content.
\begin{quote}
\textit{It all depends on how the content creator carries themselves—whether they have done the training and have the skills to create content to begin with. A lot of content creators are coming in to create content, but they have no skills. They are throwing videos out there, trying to get likes. The market is flooded with junk videos.} --P19
\end{quote}
Structured programs to guide content creators on best practices and ethical standards could significantly enhance content quality, reduce the propagation of misinformation or extremist narratives, and ultimately help maintain environments where conformity to uninformed or extreme views is less appealing.

\noindent\textbf{Interventions to prevent extremism.}
Participants discussed the importance of robust moderation strategies to proactively identify and intervene against extremist behaviors. This approach aligns closely with design implications discussed earlier for reducing conformity, such as clear guidelines, proactive moderation, and accountability mechanisms. These overlapping strategies help tackle both conformity and reinforcement by simultaneously promoting responsible discourse and discouraging extremism.
\begin{quote}
\textit{I think they can maybe have more monitors or an admin—people who can interrupt it if it is normal—but if they see that there is more violent talk, that the words they are using are a little bit more targeted, then they should be able to come in and shut it down. If it is a chat room, they should close the chat room, lock the discussion board, no more comments, and then delete it. If there are certain members who are repeat offenders, they should be banned or blocked, or have their account deleted.} --P08
\end{quote}

\subsubsection{Conditional benefits of conformity and reinforcement.}
Although participants overwhelmingly emphasized the negative consequences of conformity and reinforcement, they also described circumstances where these dynamics could serve protective or constructive purposes. These reflections underscore that the effects of silence and reinforcement are not uniformly harmful; under specific conditions, they may buffer against personal risks, preserve relationships, or amplify supportive communities.

\noindent \textbf{Benefits of conformity.} At the individual level, silence was often described as a strategic or protective choice. Withholding opinions helped some participants avoid harassment, doxxing, or stalking.  
\begin{quote}
\textit{Some part of it is security matter...You don’t want some weirdo doxxing you or stalking you. You hear about stuff like that happening all the time.} --P13
\end{quote}
Silence was also framed as a way to maintain social harmony and personal relationships, allowing individuals to navigate diverse viewpoints without unnecessary conflict.  
\begin{quote}
\textit{You don’t piss anyone off, and you don’t ruin any real life connections.} --P11
\end{quote}
\noindent \textbf{Benefits of reinforcement.}  At the community level, reinforcement within echo chambers was sometimes experienced as supportive rather than isolating. For some participants, being surrounded by like-minded voices provided a sense of belonging, validation, and confidence.  
\begin{quote}
\textit{It may foster a sense of belonging even if somebody’s view is harmful or wrong in some sense.} --P03
\end{quote}
\begin{quote}
\textit{You feel better about yourself because you think that you’re in the majority, and you think that everybody would agree with you. When you’re living in that environment, it’s good for you, or it makes your life more comfortable.} --P09
\end{quote}
Reinforcement was also seen as amplifying marginalized or contested voices, offering encouragement to persist in speaking out despite opposition.  
\begin{quote}
\textit{It allows you to express your views unfiltered, without caring about others, and usually your voice is the strongest. You influence others to start speaking up because of how unafraid you are.} --P11
\end{quote}
Taken together, these reflections show that conformity and reinforcement should not be understood in absolute terms. The same behaviors that silence diverse viewpoints in some settings can, in others, function as strategies of protection, relationship management, or community support. The critical question is not only whether people conform or reinforce, but also what consequences follow. When outcomes preserve safety, enable dialogue across differences, or empower marginalized voices, these dynamics may be adaptive. When they entrench hostility, misinformation, or exclusion, intervention becomes necessary. Recognizing this distinction shifts the discussion from simple value judgments toward designing systems that can adapt to context—across individual and community levels, and through both reactive and proactive strategies.

\section{Design Implications}

In survey and interview responses, participants highlighted critical barriers to open expression and constructive discourse on social media, emphasizing conformity pressures, fear of extremism, misinformation, and unclear moderation policies. While current platform efforts, such as fact-checking mechanisms \cite{wasike2023you} and algorithmic advancements in moderation \cite{gillespie2020content} demonstrate steps in the right direction, participants clearly indicated that these strategies alone remain insufficient to fundamentally alter user behavior or significantly reduce conformity pressures.

Participants' reflections indicate a need for a shift from reactive, moderation-focused interventions toward proactive, inclusive, and collaboration-oriented approaches. Fact-checking systems, for instance, should not operate solely as post-hoc moderation. Instead, they could be embedded into pre-posting nudges \cite{jahanbakhsh2021exploring}, for example, prompting users with counterevidence or prompting reflection when posting on high-risk or polarizing topics. This type of intervention would build on the accuracy nudges \cite{porter2024shoves,butler2024nudge} that have been studied extensively for promoting fact-checking behavior among social media users \cite{bhuiyan2021nudgecred}. By placing these nudges before posting, they may also support users who are self-censoring by providing confidence and credible evidence to express their views. Similarly, content moderation should not only penalize extreme behavior but proactively incentivize and elevate healthy, respectful interactions under community-level efforts \cite{chouhan2019co}. Platforms have experimented with peer-sourced explanatory prompts, such as Community Notes on X\footnote{https://help.x.com/en/using-x/community\_notes} (formerly known as Birdwatch on Twitter\footnote{https://blog.x.com/en\_us/topics/product/2021/introducing-birdwatch-a-community-based-approach-to-misinformation}), where trusted users provide contextual clarifications rather than just binary flags. However, these efforts often rely on voluntary participation without sustained engagement mechanisms. Our findings suggest that the effectiveness and quality of such systems could be enhanced by implementing clear incentive mechanisms—such as credibility scores, visibility boosts, or platform rewards—alongside social feedback loops like peer recognition, and by embedding them seamlessly into users' content engagement workflows.

Beyond individual-level nudges, our modeling results highlight the importance of community context in self-censorship. We found that perceived support from the community is strongly associated with whether users remain silent or adjust their expressed opinions.
This suggests opportunities for community-aware interventions. For example, platforms could enact a “small group mode” that allows sensitive posts to be shared with narrower audiences, reducing the risks associated with large, heterogeneous communities. Similarly, lightweight support signals, such as indicators that others in the network share similar views, may encourage hesitant users to participate more openly.

Importantly, study participants emphasized that relying solely on internal moderation creates a structural conflict of interest, as platforms benefit from the engagement stimulated by divisive content. To address this, a combination of internal moderation with external oversight mechanisms is essential for effectively tackling such issues \cite{elmimouni2024shielding}. One actionable approach is the implementation of independent oversight bodies, such as Meta's Oversight Board\footnote{https://www.oversightboard.com/}, which reviews content moderation decisions and provides accountability beyond the platform's internal processes. In parallel, establishing publicly accessible databases that document moderation actions would enable retrospective analysis and help restore users’ trust in actively participating in public interactions. Some platforms have taken initial steps toward this vision—for example, Twitter (now X) previously launched the Twitter Moderation Research Consortium to share moderation-related data with the public\footnote{https://blog.x.com/en\_us/topics/company/2022/twitter-moderation-research-consortium-open-researchers}. However, these efforts have not been sustained, highlighting the need for regulatory mandates to institutionalize transparency rather than relying on platform autonomy. Current regulatory efforts, including the Digital Services Act (DSA) Transparency Database\footnote{https://transparency.dsa.ec.europa.eu/}, offer a promising step toward independent auditing, though recent findings reveal uneven adherence and reporting inconsistencies across platforms \cite{trujillo2025dsa}.
These efforts underscore the importance of combining internal moderation efforts with external regulatory frameworks to strengthen platform accountability and ensure more consistent, transparent moderation practices.

Finally, we found that reinforcement and conformity coexist in a self-reinforcing loop: reinforcement intensifies conformity pressures by repeatedly amplifying dominant views and marginalizing opposite opinions, which in turn creates an environment where users feel compelled to conform even more strongly. This cyclical relationship results in progressively narrower discourse spaces, escalating polarization, and heightening the risk of online and offline harms. Effectively breaking this cycle requires targeted design interventions that address both conformity and reinforcement in tandem. These include:
\begin{itemize}
    \item Algorithmic injection of minority perspectives: Implement calibrated exposure models in feed recommendations to balance viewpoint diversity and mitigate echo chambers \cite{helberger2018exposure}.
    \item Training for responsible content creation: Provide creator education and guidelines to encourage the production of respectful and inclusive content. 
    \item Proactive, context-aware moderation: Develop moderation systems that adapt to evolving forms of interaction by considering the context of user behavior \cite{wang2024unappreciated}. 
\end{itemize}

These interventions reflect a shift toward designing for expression as a reflection of social identity, fostering community engagement and inclusivity in future platform designs.

\section{Conclusions}
Self-censorship manifests in multiple ways. Beyond complete silence, it can include posting subtle or ``toned down'' opinions, engaging passively, communicating through private channels, referencing external sources rather than commenting directly, and even adopting more extreme positions to demonstrate agreement. These patterns provide empirical support for self-silencing theory while extending it beyond complete withdrawal, showing that self-censorship often takes the form of strategic adjustment rather than silence alone. Importantly, our findings further indicate that such behaviors can reflect deliberate boundary setting or strategic restraint, rather than fear, disengagement, or withdrawal from public discourse.

While conformity effects and reinforcement through group membership might appear contradictory, they can coexist and are shaped by social context and how individuals relate to their communities. Our results refine social identity theory by showing that individuals respond not only to group membership but also to perceived community support and anticipated audience reactions. These dynamics are closely related to processes of self-presentation or ``exhibition identity'', where individuals manage how they appear to others in social settings. In these scenarios, individuals often align themselves with traits and norms deemed socially desirable. On social media, this entails highlighting aspects of one's life or opinions that are expected to attract positive feedback and validation in their community. The extent to which a person conforms or expresses their true beliefs can be influenced by their perceived community characteristics, whether they find themselves in the minority or majority within a particular social context. These patterns are also consistent with self-presentation theory and show how users actively manage the trade-off between authenticity and social acceptability. Rather than simply projecting a favorable image, participants describe ongoing negotiation between expressing true beliefs and maintaining social relationships under conditions of uncertainty and visibility.

Building on these findings, we suggest that addressing opinion conformity and reinforcement requires not only content-level interventions but also design considerations that reshape how individuals engage with identity and audience. Design implications should therefore extend beyond algorithmic adjustments or moderation strategies. They must also support more flexible self-expression and allow users to navigate complex social expectations. That is, elevating constructive discourse through fact-checking mechanisms and offering visible incentives for thoughtful engagement can help normalize meaningful conversations and reduce the pressure to conform or provoke. In addition, sustained impact requires the combined efforts of communities, platforms, and external regulatory oversight to facilitate meaningful identity presentation and enforce accountability through fair moderation practices. This shift has the potential to create more authentic, inclusive, and resilient digital environments.

\section{Limitations and Future Work}
Our findings from the survey and interviews are based on self-reported data regarding opinion sharing, which introduce biases related to self-perception. The measures in this study capture participants’ reported perceptions and intentions rather than directly observed behavior, and the results should be interpreted accordingly. This is particularly relevant for studying self-censorship, as behaviors of interest often involve non-action, such as choosing not to post, which cannot be directly observed. At the same time, self-reported data introduce their own uncertainties, and future work may benefit from combining self-reports with behavioral data where feasible.

Additionally, by limiting our participants to social media users in the United States, we excluded international users from our sample. This restriction narrows our findings to US users. While dataset statistics indicate a relatively balanced demographic distribution representing the social media user population, we acknowledge potential sampling bias. While we made efforts to reduce response bias by presenting the study materials with neutral framing and structuring interview questions to cover both positive and negative aspects, we still recognize the possibility of such bias in participants’ responses. Furthermore, while our models focused on community-level factors of self-censorship, open-ended responses revealed additional determinants such as platform norms and social influence not captured quantitatively. These factors might be studied in future work examining the complexity of self-censorship from alternative perspectives.

\section{Ethical Considerations}
Both survey and interview components of the study were approved by the Institutional Review Board (IRB) at the researchers’ institution. All participants have given their consent to participate in the survey and interview studies. All identifiable information has been removed from the paper and research artifacts to ensure privacy, and all video and audio recordings were deleted after transcription quality checks. 
Given the relatively narrow nature of the data collected in this study, we believe the risk of privacy violation and misuse to be minimal. 
\begin{acks}
This work was supported by funding from the National Science Foundation award \#2247723.
\end{acks}
\bibliographystyle{ACM-Reference-Format}
\bibliography{main.bib}

@String{Computing = "Computing" }

@String{Springer = "Springer-Verlag" }

@article{cinelli2021echo,
  title={The echo chamber effect on social media},
  author={Cinelli, Matteo and Morales, Gianmarco De Francisci and Galeazzi, Alessandro and Quattrociocchi, Walter and Starnini, Michele},
  journal={Proceedings of the National Academy of Sciences},
  volume={118},
  number={9},
  year={2021},
  publisher={National Acad Sciences}
}

@article{li2020real,
  title={Real talk about fake news: Identity language and disconnected networks of the US public’s “fake news” discourse on Twitter},
  author={Li, Jianing and Su, Min-Hsin},
  journal={Social Media+ Society},
  volume={6},
  number={2},
  pages={2056305120916841},
  year={2020},
  publisher={SAGE Publications Sage UK: London, England}
}

@article{tajfel1979integrative,
  title={An integrative theory of intergroup conflict},
  author={Tajfel, Henri and Turner, John C and Austin, William G and Worchel, Stephen},
  journal={Organizational identity: A reader},
  volume={56},
  number={65},
  pages={9780203505984--16},
  year={1979}
}

@article{jack2011reflections,
  title={Reflections on the silencing the self scale and its origins},
  author={Jack, Dana Crowley},
  journal={Psychology of Women Quarterly},
  volume={35},
  number={3},
  pages={523--529},
  year={2011},
  publisher={Sage Publications Sage CA: Los Angeles, CA}
}

@article{asch1951effects,
  title={Effects of group pressure upon the modification and distortion of judgments},
  author={Asch, Solomon E},
  journal={Organizational influence processes},
  volume={58},
  pages={295--303},
  year={1951}
}

@book{hampton2014social,
  title={Social media and the'spiral of silence'},
  author={Hampton, Keith N and Rainie, Harrison and Lu, Weixu and Dwyer, Maria and Shin, Inyoung and Purcell, Kristen},
  year={2014},
  publisher={PewResearchCenter Washington, DC, USA}
}

@article{crutchfield1955conformity,
  title={Conformity and character.},
  author={Crutchfield, Richard S},
  journal={American psychologist},
  volume={10},
  number={5},
  pages={191},
  year={1955},
  publisher={American Psychological Association}
}

@article{abbink2017peer,
  title={Peer punishment promotes enforcement of bad social norms},
  author={Abbink, Klaus and Gangadharan, Lata and Handfield, Toby and Thrasher, John},
  journal={Nature communications},
  volume={8},
  number={1},
  pages={1--8},
  year={2017},
  publisher={Nature Publishing Group}
}

@article{shang2019resilient,
  title={Resilient consensus for expressed and private opinions},
  author={Shang, Yilun},
  journal={IEEE Transactions on Cybernetics},
  volume={51},
  number={1},
  pages={318--331},
  year={2019},
  publisher={IEEE}
}

@book{allport1924social,
  title={Social psychology},
  author={Allport, Floyd Henry},
  year={1924},
  publisher={Boston, Houghton}
}

@article{centola2005emperor,
  title={The emperor’s dilemma: A computational model of self-enforcing norms},
  author={Centola, Damon and Willer, Robb and Macy, Michael},
  journal={American Journal of Sociology},
  volume={110},
  number={4},
  pages={1009--1040},
  year={2005},
  publisher={The University of Chicago Press}
}

@article{willer2009false,
  title={The false enforcement of unpopular norms},
  author={Willer, Robb and Kuwabara, Ko and Macy, Michael W},
  journal={American Journal of Sociology},
  volume={115},
  number={2},
  pages={451--490},
  year={2009},
  publisher={The University of Chicago Press}
}

@article{gastner2018consensus,
  title={Consensus time in a voter model with concealed and publicly expressed opinions},
  author={Gastner, Michael T and Oborny, Be{\'a}ta and Guly{\'a}s, M{\'a}t{\'e}},
  journal={Journal of Statistical Mechanics: Theory and Experiment},
  volume={2018},
  number={6},
  pages={063401},
  year={2018},
  publisher={IOP Publishing}
}

@article{allcott2019trends,
  title={Trends in the diffusion of misinformation on social media},
  author={Allcott, Hunt and Gentzkow, Matthew and Yu, Chuan},
  journal={Research \& Politics},
  volume={6},
  number={2},
  pages={2053168019848554},
  year={2019},
  publisher={SAGE Publications Sage UK: London, England}
}

@book{zimdars2020fake,
  title={Fake news: understanding media and misinformation in the digital age},
  author={Zimdars, Melissa and McLeod, Kembrew},
  year={2020},
  publisher={MIT Press}
}

@article{karlova2013social,
  title={A social diffusion model of misinformation and disinformation for understanding human information behaviour},
  author={Karlova, Natascha A and Fisher, Karen E},
  year={2013},
  publisher={Professor TD Wilson}
}

@article{kumar2014detecting,
  title={Detecting misinformation in online social networks using cognitive psychology},
  author={Kumar, KP Krishna and Geethakumari, G},
  journal={Human-centric Computing and Information Sciences},
  volume={4},
  number={1},
  pages={1--22},
  year={2014},
  publisher={SpringerOpen}
}

@article{del2016spreading,
  title={The spreading of misinformation online},
  author={Del Vicario, Michela and Bessi, Alessandro and Zollo, Fabiana and Petroni, Fabio and Scala, Antonio and Caldarelli, Guido and Stanley, H Eugene and Quattrociocchi, Walter},
  journal={Proceedings of the National Academy of Sciences},
  volume={113},
  number={3},
  pages={554--559},
  year={2016},
  publisher={National Acad Sciences}
}

@article{shin2018diffusion,
  title={The diffusion of misinformation on social media: Temporal pattern, message, and source},
  author={Shin, Jieun and Jian, Lian and Driscoll, Kevin and Bar, Fran{\c{c}}ois},
  journal={Computers in Human Behavior},
  volume={83},
  pages={278--287},
  year={2018},
  publisher={Elsevier}
}

@article{vosoughi2018spread,
  title={The spread of true and false news online},
  author={Vosoughi, Soroush and Roy, Deb and Aral, Sinan},
  journal={Science},
  volume={359},
  number={6380},
  pages={1146--1151},
  year={2018},
  publisher={American Association for the Advancement of Science}
}

@article{pennycook2018prior,
  title={Prior exposure increases perceived accuracy of fake news.},
  author={Pennycook, Gordon and Cannon, Tyrone D and Rand, David G},
  journal={Journal of experimental psychology: general},
  volume={147},
  number={12},
  pages={1865},
  year={2018},
  publisher={American Psychological Association}
}

@article{unkelbach2021mere,
  title={Mere Repetition Increases Belief in Factually True COVID-19-Related Information},
  author={Unkelbach, Christian and Speckmann, Felix},
  journal={Journal of Applied Research in Memory and Cognition},
  year={2021},
  publisher={Elsevier}
}

@article{douglas2023data,
  title={Data quality in online human-subjects research: Comparisons between MTurk, Prolific, CloudResearch, Qualtrics, and SONA},
  author={Douglas, Benjamin D and Ewell, Patrick J and Brauer, Markus},
  journal={Plos one},
  volume={18},
  number={3},
  pages={e0279720},
  year={2023},
  publisher={Public Library of Science San Francisco, CA USA}
}

@article{oppenheimer2009instructional,
  title={Instructional manipulation checks: Detecting satisficing to increase statistical power},
  author={Oppenheimer, Daniel M and Meyvis, Tom and Davidenko, Nicolas},
  journal={Journal of experimental social psychology},
  volume={45},
  number={4},
  pages={867--872},
  year={2009},
  publisher={Elsevier}
}

@article{king1981conflicts,
  title={Conflicts between public and private opinion},
  author={King, Stephen},
  journal={Long Range Planning},
  volume={14},
  number={4},
  pages={90--105},
  year={1981},
  publisher={Elsevier}
}

@article{rose2007going,
  title={Going Public with Private Opinions: Are Post-Communist Citizens Afraid to Say What They Think?},
  author={Rose, Richard},
  journal={Journal of Elections, Public Opinion and Parties},
  volume={17},
  number={2},
  pages={123--142},
  year={2007},
  publisher={Taylor \& Francis}
}

@incollection{baumeister1987self,
  title={Self-presentation theory: Self-construction and audience pleasing},
  author={Baumeister, Roy F and Hutton, Debra G},
  booktitle={Theories of group behavior},
  pages={71--87},
  year={1987},
  publisher={Springer}
}

@article{goffman1949presentation,
  title={Presentation of self in everyday life},
  author={Goffman, Erving},
  journal={American Journal of Sociology},
  volume={55},
  pages={6--7},
  year={1949}
}

@article{schlenker1985identity,
  title={Identity and self-identification},
  author={Schlenker, Barry R},
  journal={The self and social life},
  volume={65},
  number={1},
  pages={99--106},
  year={1985}
}

@article{hollenbaugh2021self,
  title={Self-presentation in social media: Review and research opportunities},
  author={Hollenbaugh, Erin E},
  journal={Review of communication research},
  volume={9},
  pages={80},
  year={2021},
  publisher={Review of Communication Research}
}

@article{hennink2022sample,
  title={Sample sizes for saturation in qualitative research: A systematic review of empirical tests},
  author={Hennink, Monique and Kaiser, Bonnie N},
  journal={Social science \& medicine},
  volume={292},
  pages={114523},
  year={2022},
  publisher={Elsevier}
}

@article{tajfel2004social,
  title={The social identity theory of intergroup behavior.},
  author={Tajfel, Henri and Turner, John C},
  year={2004},
  publisher={Psychology Press}
}

@article{stets2000identity,
  title={Identity theory and social identity theory},
  author={Stets, Jan E and Burke, Peter J},
  journal={Social psychology quarterly},
  pages={224--237},
  year={2000},
  publisher={JSTOR}
}

@article{brewer1999psychology,
  title={The psychology of prejudice: Ingroup love or outgroup hate?},
  author={Brewer, Marilynn B and others},
  journal={Journal of social issues},
  volume={55},
  pages={429--444},
  year={1999},
  publisher={Citeseer}
}

@article{greene2004social,
  title={Social identity theory and party identification},
  author={Greene, Steven},
  journal={Social Science Quarterly},
  volume={85},
  number={1},
  pages={136--153},
  year={2004},
  publisher={Wiley Online Library}
}

@article{trepte2017social,
  title={Social identity theory and self-categorization theory},
  author={Trepte, Sabine and Loy, Laura S},
  journal={The international encyclopedia of media effects},
  pages={1--13},
  year={2017},
  publisher={John Wiley \& Sons, Inc. Hoboken, NJ, USA}
}

@inproceedings{bai2019exploring,
  title={Exploring cognitive dissonance on social media},
  author={Bai, Jie and Kong, Qingchao and Li, Linjing and Wang, Lei and Zeng, Daniel},
  booktitle={2019 IEEE International Conference on Intelligence and Security Informatics (ISI)},
  pages={143--145},
  year={2019},
  organization={IEEE}
}

@misc{ganesh2020countering,
  title={Countering extremists on social media: Challenges for strategic communication and content moderation},
  author={Ganesh, Bharath and Bright, Jonathan},
  journal={Policy \& Internet},
  volume={12},
  number={1},
  pages={6--19},
  year={2020},
  publisher={Wiley Online Library}
}

@article{clifford2021moderating,
  title={Moderating Extremism: The State of Online Terrorist Content Removal Policy in the United States},
  author={Clifford, Bennett},
  journal={Program on Extremism. Washington DC: George Washington University, December},
  year={2021}
}

@article{lakomy2023online,
  title={Why do online countering violent extremism strategies not work? The case of digital jihad},
  author={Lakomy, Miron},
  journal={Terrorism and political violence},
  volume={35},
  number={6},
  pages={1261--1298},
  year={2023},
  publisher={Taylor \& Francis}
}

@article{gearhart2015something,
  title={“Was it something I said?”“No, it was something you posted!” A study of the spiral of silence theory in social media contexts},
  author={Gearhart, Sherice and Zhang, Weiwu},
  journal={Cyberpsychology, Behavior, and Social Networking},
  volume={18},
  number={4},
  pages={208--213},
  year={2015},
  publisher={Mary Ann Liebert, Inc. 140 Huguenot Street, 3rd Floor New Rochelle, NY 10801 USA}
}

@article{sohn2022spiral,
  title={Spiral of silence in the social media era: A simulation approach to the interplay between social networks and mass media},
  author={Sohn, Dongyoung},
  journal={Communication Research},
  volume={49},
  number={1},
  pages={139--166},
  year={2022},
  publisher={SAGE Publications Sage CA: Los Angeles, CA}
}

@article{chaudhry2020expressing,
  title={Expressing and challenging racist discourse on Facebook: How social media weaken the “spiral of silence” theory},
  author={Chaudhry, Irfan and Gruzd, Anatoliy},
  journal={Policy \& Internet},
  volume={12},
  number={1},
  pages={88--108},
  year={2020},
  publisher={Wiley Online Library}
}

@inproceedings{ballard2022conspiracy,
  title={Conspiracy brokers: Understanding the monetization of youtube conspiracy theories},
  author={Ballard, Cameron and Goldstein, Ian and Mehta, Pulak and Smothers, Genesis and Take, Kejsi and Zhong, Victoria and Greenstadt, Rachel and Lauinger, Tobias and McCoy, Damon},
  booktitle={Proceedings of the ACM Web Conference 2022},
  pages={2707--2718},
  year={2022}
}

@article{zar2005spearman,
  title={Spearman rank correlation},
  author={Zar, Jerrold H},
  journal={Encyclopedia of biostatistics},
  volume={7},
  year={2005},
  publisher={Wiley Online Library}
}

@article{nick2007logistic,
  title={Logistic regression},
  author={Nick, Todd G and Campbell, Kathleen M},
  journal={Topics in biostatistics},
  pages={273--301},
  year={2007},
  publisher={Springer}
}

@article{fullerton2009conceptual,
  title={A conceptual framework for ordered logistic regression models},
  author={Fullerton, Andrew S},
  journal={Sociological methods \& research},
  volume={38},
  number={2},
  pages={306--347},
  year={2009},
  publisher={Sage Publications Sage CA: Los Angeles, CA}
}

@article{mackinnon2007mediation,
  title={Mediation analysis},
  author={MacKinnon, David P and Fairchild, Amanda J and Fritz, Matthew S},
  journal={Annu. Rev. Psychol.},
  volume={58},
  pages={593--614},
  year={2007},
  publisher={Annual Reviews}
}

@article{braun2006using,
  title={Using thematic analysis in psychology},
  author={Braun, Virginia and Clarke, Victoria},
  journal={Qualitative research in psychology},
  volume={3},
  number={2},
  pages={77--101},
  year={2006},
  publisher={Taylor \& Francis}
}

@article{dhakal2022nvivo,
  title={NVivo},
  author={Dhakal, Kerry},
  journal={Journal of the Medical Library Association: JMLA},
  volume={110},
  number={2},
  pages={270},
  year={2022},
  publisher={Medical Library Association}
}

@inproceedings{wu2024reacting,
author = {Wu, Chuhao and Wang, Xinyu and Carroll, John and Rajtmajer, Sarah},
title = {Reacting to Generative AI: Insights from Student and Faculty Discussions on Reddit},
year = {2024},
isbn = {9798400703348},
publisher = {Association for Computing Machinery},
address = {New York, NY, USA},
url = {https://doi.org/10.1145/3614419.3644014},
doi = {10.1145/3614419.3644014},
booktitle = {Proceedings of the 16th ACM Web Science Conference},
pages = {103–113},
numpages = {11},
location = {, Stuttgart, Germany, },
series = {WEBSCI '24}
}

@inproceedings{ammari2015understanding,
  title={Understanding and supporting fathers and fatherhood on social media sites},
  author={Ammari, Tawfiq and Schoenebeck, Sarita},
  booktitle={Proceedings of the 33rd annual ACM conference on human factors in computing systems},
  pages={1905--1914},
  year={2015}
}

@article{lee2021digital,
  title={Digital inequality through the lens of self-disclosure},
  author={Lee, Jooyoung and Rajtmajer, Sarah and Srivatsavaya, Eesha and Wilson, Shomir},
  journal={Proceedings on Privacy Enhancing Technologies},
  year={2021}
}

@article{eyal2021data,
  title={Data quality of platforms and panels for online behavioral research},
  author={Eyal, Peer and David, Rothschild and Andrew, Gordon and Zak, Evernden and Ekaterina, Damer},
  journal={Behavior research methods},
  pages={1--20},
  year={2021},
  publisher={Springer}
}

@article{lim2022opinion,
  title={Opinion amplification causes extreme polarization in social networks},
  author={Lim, Soo Ling and Bentley, Peter J},
  journal={Scientific Reports},
  volume={12},
  number={1},
  pages={18131},
  year={2022},
  publisher={Nature Publishing Group UK London}
}

@article{konovalova2023social,
  title={Social media feedback and extreme opinion expression},
  author={Konovalova, Elizaveta and Le Mens, Ga{\"e}l and Sch{\"o}ll, Nikolas},
  journal={Plos one},
  volume={18},
  number={11},
  pages={e0293805},
  year={2023},
  publisher={Public Library of Science San Francisco, CA USA}
}

@article{neureiter2021trust,
  title={Trust in science, perceived media exaggeration about COVID-19, and social distancing behavior},
  author={Neureiter, Ariadne and Stubenvoll, Marlis and Kaskeleviciute, Ruta and Matthes, J{\"o}rg},
  journal={Frontiers in Public Health},
  volume={9},
  pages={670485},
  year={2021},
  publisher={Frontiers Media SA}
}

@article{oz2024platform,
  title={Platform affordances and spiral of silence: How perceived differences between Facebook and Twitter influence opinion expression online},
  author={Oz, Mustafa and Shahin, Saif and Greeves, Scott B},
  journal={Technology in Society},
  volume={76},
  pages={102431},
  year={2024},
  publisher={Elsevier}
}

@article{clayton2020real,
  title={Real solutions for fake news? Measuring the effectiveness of general warnings and fact-check tags in reducing belief in false stories on social media},
  author={Clayton, Katherine and Blair, Spencer and Busam, Jonathan A and Forstner, Samuel and Glance, John and Green, Guy and Kawata, Anna and Kovvuri, Akhila and Martin, Jonathan and Morgan, Evan and others},
  journal={Political behavior},
  volume={42},
  pages={1073--1095},
  year={2020},
  publisher={Springer}
}

@article{walter2020fact,
  title={Fact-checking: A meta-analysis of what works and for whom},
  author={Walter, Nathan and Cohen, Jonathan and Holbert, R Lance and Morag, Yasmin},
  journal={Political communication},
  volume={37},
  number={3},
  pages={350--375},
  year={2020},
  publisher={Taylor \& Francis}
}

@article{racca2018relating,
  title={Relating group size and posting activity of an online community of financial investors: Regularities and seasonal patterns},
  author={Racca, P and Casarin, Roberto and Dondio, Pierpaolo and Squazzoni, F},
  journal={Physica A: Statistical Mechanics and its Applications},
  volume={493},
  pages={458--466},
  year={2018},
  publisher={Elsevier}
}

@article{wasike2023you,
  title={You've been fact-checked! Examining the effectiveness of social media fact-checking against the spread of misinformation},
  author={Wasike, Ben},
  journal={Telematics and Informatics Reports},
  volume={11},
  pages={100090},
  year={2023},
  publisher={Elsevier}
}

@article{gillespie2020content,
  title={Content moderation, AI, and the question of scale},
  author={Gillespie, Tarleton},
  journal={Big Data \& Society},
  volume={7},
  number={2},
  pages={2053951720943234},
  year={2020},
  publisher={SAGE Publications Sage UK: London, England}
}

@article{elmimouni2024shielding,
  title={Shielding or silencing?: An investigation into content moderation during the sheikh jarrah crisis},
  author={Elmimouni, Houda and Skop, Yarden and Abokhodair, Norah and R{\"u}ller, Sarah and Aal, Konstantin and Weibert, Anne and Al-Dawood, Adel and Wulf, Volker and Tolmie, Peter},
  journal={Proceedings of the ACM on Human-Computer Interaction},
  volume={8},
  number={GROUP},
  pages={1--21},
  year={2024},
  publisher={ACM New York, NY, USA}
}

@article{chouhan2019co,
  title={Co-designing for community oversight: Helping people make privacy and security decisions together},
  author={Chouhan, Chhaya and LaPerriere, Christy M and Aljallad, Zaina and Kropczynski, Jess and Lipford, Heather and Wisniewski, Pamela J},
  journal={Proceedings of the ACM on Human-Computer Interaction},
  volume={3},
  number={CSCW},
  pages={1--31},
  year={2019},
  publisher={ACM New York, NY, USA}
}

@article{wang2024unappreciated,
  title={The unappreciated role of intent in algorithmic moderation of social media content},
  author={Wang, Xinyu and Koneru, Sai and Venkit, Pranav Narayanan and Frischmann, Brett and Rajtmajer, Sarah},
  journal={arXiv preprint arXiv:2405.11030},
  year={2024}
}

@article{helberger2018exposure,
  title={Exposure diversity as a design principle for recommender systems},
  author={Helberger, Natali and Karppinen, Kari and D’acunto, Lucia},
  journal={Information, communication \& society},
  volume={21},
  number={2},
  pages={191--207},
  year={2018},
  publisher={Taylor \& Francis}
}

@article{jahanbakhsh2021exploring,
  title={Exploring lightweight interventions at posting time to reduce the sharing of misinformation on social media},
  author={Jahanbakhsh, Farnaz and Zhang, Amy X and Berinsky, Adam J and Pennycook, Gordon and Rand, David G and Karger, David R},
  journal={Proceedings of the ACM on human-computer interaction},
  volume={5},
  number={CSCW1},
  pages={1--42},
  year={2021},
  publisher={ACM New York, NY, USA}
}

@article{bhuiyan2021nudgecred,
  title={Nudgecred: Supporting news credibility assessment on social media through nudges},
  author={Bhuiyan, Md Momen and Horning, Michael and Lee, Sang Won and Mitra, Tanushree},
  journal={Proceedings of the ACM on Human-Computer Interaction},
  volume={5},
  number={CSCW2},
  pages={1--30},
  year={2021},
  publisher={ACM New York, NY, USA}
}

@article{porter2024shoves,
  title={Shoves, nudges and combating misinformation: evidence on a new approach},
  author={Porter, Ethan and Wood, Thomas J and Broniatowski, David A and Hosseini, Pedram},
  journal={Behavioural Public Policy},
  pages={1--17},
  year={2024},
  publisher={Cambridge University Press}
}

@article{butler2024nudge,
  title={Nudge-based misinformation interventions are effective in information environments with low misinformation prevalence},
  author={Butler, Lucy H and Prike, Toby and Ecker, Ullrich KH},
  journal={Scientific Reports},
  volume={14},
  number={1},
  pages={11495},
  year={2024},
  publisher={Nature Publishing Group UK London}
}

@article{barbera2020social,
  title={Social media, echo chambers, and political polarization},
  author={Barber{\'a}, Pablo},
  journal={Social media and democracy: The state of the field, prospects for reform},
  pages={34--55},
  year={2020},
  publisher={SSRC Anxieties of Democracy, Cambridge University Press, Cambridge, UK}
}

@article{rhodes2022filter,
  title={Filter bubbles, echo chambers, and fake news: How social media conditions individuals to be less critical of political misinformation},
  author={Rhodes, Samuel C},
  journal={Political Communication},
  volume={39},
  number={1},
  pages={1--22},
  year={2022},
  publisher={Taylor \& Francis}
}

@article{barbera2015tweeting,
  title={Tweeting from left to right: Is online political communication more than an echo chamber?},
  author={Barber{\'a}, Pablo and Jost, John T and Nagler, Jonathan and Tucker, Joshua A and Bonneau, Richard},
  journal={Psychological science},
  volume={26},
  number={10},
  pages={1531--1542},
  year={2015},
  publisher={Sage Publications Sage CA: Los Angeles, CA}
}

@inproceedings{bruns2017echo,
  title={Echo chamber? What echo chamber? Reviewing the evidence},
  author={Bruns, Axel},
  booktitle={6th Biennial Future of Journalism Conference (FOJ17)},
  year={2017}
}

@article{guess2018avoiding,
  title={Avoiding the echo chamber about echo chambers},
  author={Guess, Andrew and Nyhan, Brendan and Lyons, Benjamin and Reifler, Jason},
  journal={Knight Foundation},
  volume={2},
  number={1},
  pages={1--25},
  year={2018}
}

@inproceedings{garimella2018political,
  title={Political discourse on social media: Echo chambers, gatekeepers, and the price of bipartisanship},
  author={Garimella, Kiran and De Francisci Morales, Gianmarco and Gionis, Aristides and Mathioudakis, Michael},
  booktitle={Proceedings of the 2018 world wide web conference},
  pages={913--922},
  year={2018}
}

@inproceedings{nguyen2014exploring,
  title={Exploring the filter bubble: the effect of using recommender systems on content diversity},
  author={Nguyen, Tien T and Hui, Pik-Mai and Harper, F Maxwell and Terveen, Loren and Konstan, Joseph A},
  booktitle={Proceedings of the 23rd international conference on World wide web},
  pages={677--686},
  year={2014}
}

@article{bozdag2013bias,
  title={Bias in algorithmic filtering and personalization},
  author={Bozdag, Engin},
  journal={Ethics and information technology},
  volume={15},
  pages={209--227},
  year={2013},
  publisher={Springer}
}

@article{bail2018exposure,
  title={Exposure to opposing views on social media can increase political polarization},
  author={Bail, Christopher A and Argyle, Lisa P and Brown, Taylor W and Bumpus, John P and Chen, Haohan and Hunzaker, MB Fallin and Lee, Jaemin and Mann, Marcus and Merhout, Friedolin and Volfovsky, Alexander},
  journal={Proceedings of the National Academy of Sciences},
  volume={115},
  number={37},
  pages={9216--9221},
  year={2018},
  publisher={National Academy of Sciences}
}

@article{hobolt2024polarizing,
  title={The polarizing effect of partisan echo chambers},
  author={Hobolt, Sara B and Lawall, Katharina and Tilley, James},
  journal={American Political Science Review},
  volume={118},
  number={3},
  pages={1464--1479},
  year={2024},
  publisher={Cambridge University Press}
}

@inproceedings{gillani2018me,
  title={Me, my echo chamber, and I: introspection on social media polarization},
  author={Gillani, Nabeel and Yuan, Ann and Saveski, Martin and Vosoughi, Soroush and Roy, Deb},
  booktitle={Proceedings of the 2018 World Wide Web Conference},
  pages={823--831},
  year={2018}
}

@article{warner2019self,
  title={Self-censorship in social networking sites (SNSs)--privacy concerns, privacy awareness, perceived vulnerability and information management},
  author={Warner, Mark and Wang, Victoria},
  journal={Journal of Information, Communication and Ethics in Society},
  volume={17},
  number={4},
  pages={375--394},
  year={2019},
  publisher={Emerald Publishing Limited}
}

@article{gibson2023keeping,
  title={Keeping your mouth shut: Spiraling self-censorship in the United States},
  author={Gibson, James L and Sutherland, Joseph L},
  journal={Political Science Quarterly},
  volume={138},
  number={3},
  pages={361--376},
  year={2023},
  publisher={Oxford University Press US}
}

@inproceedings{howe2023self,
  title={Self-censorship appears to be an effective way of reducing the spread of misinformation on social media},
  author={Howe, Piers and Perfors, Andrew and Ransom, Keith James and Walker, Bradley and Fay, Nicolas and Kashima, Yoshihisa and Saletta, Morgan},
  booktitle={Proceedings of the Annual Meeting of the Cognitive Science Society},
  volume={45},
  number={45},
  year={2023}
}

@inproceedings{dubois2018self,
  title={Self-Censorship, Polarization, and the―Spiral of Silence on Social Media},
  author={Dubois, Elizabeth and Szwarc, Julia},
  booktitle={Policy \& Politics Conference},
  year={2018}
}

@article{kwon2015unspeaking,
  title={Unspeaking on Facebook? Testing network effects on self-censorship of political expressions in social network sites},
  author={Kwon, K Hazel and Moon, Shin-Il and Stefanone, Michael A},
  journal={Quality \& quantity},
  volume={49},
  pages={1417--1435},
  year={2015},
  publisher={Springer}
}

@article{coleman1988social,
  title={Social capital in the creation of human capital},
  author={Coleman, James S},
  journal={American journal of sociology},
  volume={94},
  pages={S95--S120},
  year={1988},
  publisher={University of Chicago Press}
}

@article{brandtzaeg2010too,
  title={Too many Facebook “friends”? Content sharing and sociability versus the need for privacy in social network sites},
  author={Brandtz{\ae}g, Petter Bae and L{\"u}ders, Marika and Skjetne, Jan H{\aa}vard},
  journal={Intl. Journal of Human--Computer Interaction},
  volume={26},
  number={11-12},
  pages={1006--1030},
  year={2010},
  publisher={Taylor \& Francis}
}

@inproceedings{das2013self,
  title={Self-censorship on Facebook},
  author={Das, Sauvik and Kramer, Adam},
  booktitle={Proceedings of the International AAAI Conference on Web and Social Media},
  volume={7},
  number={1},
  pages={120--127},
  year={2013}
}

@article{zhao2025mapping,
  title={Mapping the Spiral of Silence: Surveying Unspoken Opinions in Online Communities},
  author={Zhao, Dora and Yang, Diyi and Bernstein, Michael S},
  journal={arXiv preprint arXiv:2502.00952},
  year={2025}
}

@article{filak2009expanding,
  title={Expanding and validating applications of the willingness to self-censor scale: Self-censorship and media advisers' comfort level with controversial topics},
  author={Filak, Vincent F and Reinardy, Scott and Maksl, Adam},
  journal={Journalism \& Mass Communication Quarterly},
  volume={86},
  number={2},
  pages={368--382},
  year={2009},
  publisher={SAGE Publications Sage CA: Los Angeles, CA}
}

@article{burnett2022self,
  title={The self-censoring majority: How political identity and ideology impacts willingness to self-censor and fear of isolation in the United States},
  author={Burnett, Alycia and Knighton, Devin and Wilson, Christopher},
  journal={Social Media+ Society},
  volume={8},
  number={3},
  pages={20563051221123031},
  year={2022},
  publisher={SAGE Publications Sage UK: London, England}
}

@article{bar2017self,
  title={Self-censorship as a socio-political-psychological phenomenon: Conception and research},
  author={Bar-Tal, Daniel},
  journal={Political Psychology},
  volume={38},
  pages={37--65},
  year={2017},
  publisher={Wiley Online Library}
}

@article{trujillo2025dsa,
  title={The DSA Transparency Database: Auditing self-reported moderation actions by social media},
  author={Trujillo, Amaury and Fagni, Tiziano and Cresci, Stefano},
  journal={Proceedings of the ACM on Human-Computer Interaction},
  volume={9},
  number={2},
  pages={1--28},
  year={2025},
  publisher={ACM New York, NY, USA}
}

@inproceedings{gera2020hesitation,
  title={Hesitation while posting: A cross-sectional survey of sensitive topics and opinion sharing on social media},
  author={Gera, Parush and Thomas, Nadia and Neal, Tempestt},
  booktitle={International Conference on Social Media and Society},
  pages={134--140},
  year={2020}
}

@article{weeks2024too,
  title={Too scared to share? Fear of social sanctions for political expression on social media},
  author={Weeks, Brian E and Halversen, Audrey and Neubaum, German},
  journal={Journal of Computer-Mediated Communication},
  volume={29},
  number={1},
  pages={zmad041},
  year={2024},
  publisher={Oxford University Press}
}

@article{powers2019shouting,
  title={“Shouting matches and echo chambers”: perceived identity threats and political self-censorship on social media},
  author={Powers, Elia and Koliska, Michael and Guha, Pallavi},
  journal={International Journal of Communication},
  volume={13},
  pages={20},
  year={2019}
}

@article{juncosa2024toxic,
  title={Toxic behavior silences online political conversations},
  author={Juncosa, Gabriela and Yasseri, Taha and Koltai, Julia and Iniguez, Gerardo},
  journal={arXiv preprint arXiv:2412.05741},
  year={2024}
}

@article{palinkas2015purposeful,
  title={Purposeful sampling for qualitative data collection and analysis in mixed method implementation research},
  author={Palinkas, Lawrence A and Horwitz, Sarah M and Green, Carla A and Wisdom, Jennifer P and Duan, Naihua and Hoagwood, Kimberly},
  journal={Administration and policy in mental health and mental health services research},
  volume={42},
  number={5},
  pages={533--544},
  year={2015},
  publisher={Springer}
}

@book{patton2014qualitative,
  title={Qualitative research \& evaluation methods: Integrating theory and practice},
  author={Patton, Michael Quinn},
  year={2014},
  publisher={Sage publications}
}

@article{noelle1974spiral,
  title={The spiral of silence a theory of public opinion},
  author={Noelle-Neumann, Elisabeth},
  journal={Journal of communication},
  volume={24},
  number={2},
  pages={43--51},
  year={1974},
  publisher={Oxford University Press}
}

@article{marwick2011tweet,
  title={I tweet honestly, I tweet passionately: Twitter users, context collapse, and the imagined audience},
  author={Marwick, Alice E and Boyd, Danah},
  journal={New media \& society},
  volume={13},
  number={1},
  pages={114--133},
  year={2011},
  publisher={Sage Publications Sage UK: London, England}
}

\appendix

\renewcommand{\thetable}{A\arabic{table}}
\setcounter{table}{0} 
\section{Appendix 1: Survey Instrument}
\subsection{Introduction}
\textbf{Objectives}: This survey aims to understand patterns of opinion expression on social media platforms. Your responses are anonymous and will help understand the dynamics of online discourse.\\
\textbf{Privacy and Ethics Statement}: The survey data collected is used solely for research purposes and all responses are kept strictly confidential. Respondents have the right to withdraw at any time without penalty or retribution. By participating, respondents consent to the use of their anonymized data for research purposes.

\subsection{Consent}
Please consent to the following statement: I agree to participate in the research study. I understand the purpose and nature of this study and I am participating voluntarily. I understand that I can withdraw from the study at any time, without any penalty or consequences. The IRB has reviewed the study and determined that, as currently described, it is exempt from ongoing IRB review.

\subsection{Demographic Information}

1. What is your gender?\\
2. What category below includes your age?\\
3. Which race/ethnicity best describes you?\\
4. What is the highest degree you have completed?\\
5. In politics today, do you consider yourself a Republican, Democrat, an independent or something else?\\
(If answer ``Independent or other'': As of today, do you lean more to the Republican Party or more to the Democratic Party?)
\subsection{Social Media Usage}
\noindent 1. Which social media platforms do you use regularly and how many connections do you have on the target platforms?\\
2. In a typical week, how many times do you share or post your opinions on social media platforms related to one or more of the following topics: politics (e.g., US election), social justice issues (e.g., gender equality), environmental issues (e.g., climate change), religion (e.g., ritual practices), health and healthcare (e.g., vaccinations), or technology and innovation (e.g., ChatGPT)?

\subsection{Opinion Expression}
\noindent 1. Assess the typical/average level of polarization you observe in conversations online about each of the following topics in your social networks.\\
2. Are you more or less willing to express your opinions about the following topics on social media?\\
\noindent \textit{3. Please indicate the extent to which you agree with each of the following statements:}\\
I feel more authentic in expressing my beliefs in private conversations than on social media platforms.\\
I feel the need to filter or adjust my personal beliefs before sharing them on social media.\\
Sometimes my social media persona is different from my real-life persona in terms of expressing opinions.\\
\textit{4. Please indicate the extent to which you agree with each of the following statements:}\\
I have censored or modified my opinion before posting due to fear of isolation or criticism.\\
I have refrained from expressing a minority view on a polarizing topic to avoid confrontation.\\
I feel that certain views are silenced on social media because they are unpopular.\\
\textit{5. Please indicate the extent to which you agree with each of the following statements:}\\
I feel more inclined to express controversial opinions on social media because I anticipate support from my social media connections.\\
I feel that my opinions are reinforced by my social media circle.\\
I believe that the backing of my in-group on social media leads me to amplify my expressed opinions.

\subsection{Topic-specific Questions}
\textbf{1. Politics}\\
Which side are you leaning towards in the upcoming U.S. election?\\
In your opinion, how polarizing is this topic of the upcoming U.S. election?\\
How many of your connections do you estimate share the same viewpoint as you on the upcoming U.S. election? (expressed as a percentage)\\
Based on your current viewpoint, which one of the following posts would you be more inclined to share?\\
\textbf{2. Health}\\
What is your opinion on COVID-19 vaccination?\\
In your opinion, how polarizing is the topic of COVID-19 vaccination?\\
How many of your connections do you estimate share the same viewpoint as you on COVID-19 vaccination? (expressed as a percentage)\\
Based on your current viewpoint, which one of the following posts would you be more inclined to share?\\
\textbf{3. Technology}\\
What is your opinion on ChatGPT?\\
In your opinion, how polarizing is this topic of ChatGPT?\\
How many of your connections do you estimate share the same viewpoint as you on ChatGPT? (expressed as a percentage)\\
Based on your current viewpoint, which one of the following posts would you be more inclined to share?\\
\textbf{4. Social Justice}\\
What is your opinion on immigrants/immigration?\\
In your opinion, how polarizing is the topic of immigrants/immigration?\\
How many of your connections do you estimate share the same viewpoint as you on immigrants/immigration? (expressed as a percentage)\\
Based on your current viewpoint, which opinion would you be more inclined to share?\\
\textbf{5. Religion}\\
What is your opinion regarding the intersection of traditional religious beliefs and support for LGBT rights?\\
In your opinion, how polarizing is this topic of the intersection of traditional religious beliefs and support for LGBT rights?\\
How many of your connections do you estimate share the same viewpoint as you on the intersection of traditional religious beliefs and support for LGBT rights? (expressed as a percentage)\\
Based on your current viewpoint, which one of the following posts would you be more inclined to share?\\
\textbf{6. Environment}\\
What is your opinion regarding the trade-off between economic development and the urgency of climate action?\\
In your opinion, how polarizing is this topic of the trade-off between economic development and the urgency of climate action?\\
How many of your connections do you estimate share the same viewpoint as you on trade-off between economic development and the urgency of climate action? (expressed as a percentage)\\
Based on your current viewpoint, which one of the following posts would you be more inclined to share?
\subsection{Open-ended Questions}
1. Topics of Disparity: Please specify a topic on which you've noticed a disparity between your own social media posts and your private beliefs.\\
2. Recent Social Media Post: Provide the content (or a summary) of one of your recent posts on this topic.\\
3. Your Private Belief: Describe your actual belief or opinion on this topic.
4. Reason for Disparity: If there's a difference between your recent posts and your true belief, could you explain why?\\
5. Frequency of Disparity: How often do you find that your posts on social media differ from your private beliefs across various topics?\\
6. Additional Topics: Are there any other topics or subjects where you've noticed a similar disparity between your social media posts and your private beliefs? If so, please elaborate.\\
7. Potential Impact: Do you think reluctance to express true opinions could potentially lead to negative consequences? If so, what could be the consequences? If not, why?

\subsection{Invitation to Participate in the Interview}
Thank you for participating!\\
If you would like to be considered for a paid post-survey study (\$15 for 30-minute interview), please reach out to us for more information.

\section{Appendix 2: Interview Questions}
1. Discrepancy Between Public and Private Opinions\\
a. Instances of Discrepancy\\
•	Have you ever expressed an opinion on social media that differed from your private opinion?\\
•	Please specify the topic and outline what you said versus your actual opinion.\\
•	What prompted you to present a different opinion publicly?\\
b. Observation of Others' Discrepancies\\
•	If you seldom post, have you noticed others displaying a discrepancy between their authentic and public opinions on social media?\\
c. Consistency in Opinions\\
•	If you always keep your public and private opinions the same, explain why.\\
2. Determinants of Opinion Sharing\\
•	How do you determine which opinions to share or not share on social media?\\
•	How does the topic of discussion influence your willingness to share certain opinions publicly? \\
•	What features of a topic make you more willing or unwilling to share your opinion?\\
3. Self-presentation through Social Media\\
•	Do you consciously curate a specific image on social media with the way you post (either post or not post)?\\
•	Do you feel any conflict between presenting an image of yourself on social media platforms and being your true self?\\
4. Influence of Social Media Community\\
•	How does the level of support from your connections affect your willingness to share opinions?\\
5. The Consequences and Influence of Social Media Dynamics\\
•	Do you witness more opinion reinforcement/amplification or conformity on social media?\\
•	If observe more reinforcement/amplification: What do you consider to be the potential positive and negative outcomes of opinion reinforcement and amplification?\\
•	If observe more conformity: What could be the potential positive and negative consequences of opinion conformity and the spiral of silence?\\
•	Do you think the social media community or platform should address these issues?
\section{Appendix 3}
\subsection{Quality Check Procedure}
At the beginning of the survey we asked participants about their political orientation. Then, approximately midway through the survey (within a topic-specific section on Politics), we asked participants a similar question once again. This attention check was seamlessly integrated into the flow of the survey. We identified 1 out of 390 responses which exhibited a contradictory response on the attention check. We contacted that participant for an explanation and re-evaluation. The participant explained their reasoning, thus their survey was included in downstream analyses. Eleven (11) out of 390 responses were flagged by Qualtrics as potential bots. We undertook meticulous examination of those flagged answers including the completeness of the open-ended questions and total time spent on the survey. We manually ruled out the possibility of those responses being bots. Thus, we have 390 valid survey responses that are analyzed in the paper.

\begin{table*}[ht]
\centering
\small
\begin{tabular}{@{}lc@{}}
\toprule
\textbf{Characteristic}                      & \textbf{Values, \( n (\%) \)} \\ \midrule
\textbf{Gender}              \\
  Female                         &   204, 52.31\%       \\
  Male                  &   172, 44.10\%       \\
  Other or Prefer not to disclose     &  14,  3.59\%         \\ \addlinespace
\textbf{Age Group}               \\
18 to 24        & 53, 13.59\%          \\
25 to 34  & 128, 32.82\%        \\
35 to 44 &   92,23.59\%         \\
45 to 54           &  65, 16.67\%   \\ 
Above 55           &   52, 13.33\%       \\ \addlinespace
\textbf{Race/Ethnicity}               \\
White/Caucasian                           &     218, 55.9\%\\
Hispanic, Latino, or of Spanish origin                  &   50, 12.82\%      \\
Black/African American                       &    46, 11.79\%       \\
Asian           &   49,    12.56\%     \\ 
Native American           &    4, 1.03\%       \\ 
Multiple Ethnicity  &   23, 5.90\%        \\ \addlinespace
\textbf{Highest degree of education completed}         \\
Less than high school degree&  6, 1.54\%                 \\
High school degree or equivalent            & 98, 25.13\%         \\
Attending College                       &  39, 10.00\%       \\
Associate degree                        &  42, 10.77\%       \\
Bachelor degree                      &  146, 37.44\%     \\
Graduate degree                        &   51, 13.08\%       \\ 
Other                        & 8, 2.05\%         \\ \addlinespace
\textbf{Political Self-identification} &          \\
Republican                  &   69, 17.69\%    \\
Democrat              & 195, 50.00\%       \\
Independent or others (Leaning Democrat)        &  77, 19.74\%   \\
Independent or others (Leaning Republican)       &  47, 12.05\%      \\
Independent or others (No Lean/Neutral)           &  2, 0.51\%     \\
\bottomrule
\end{tabular}
\caption{Survey respondent demographics (N=390).}
\label{tab:statistics}
\end{table*}

\begin{table*}[h]
\centering
\setlength{\tabcolsep}{2.7pt} 
\renewcommand{\arraystretch}{1} 
\begin{tabular}{@{}lcccc@{}} 
\toprule
 & \textbf{Support} & \textbf{Polarization} & \textbf{Community} & \textbf{Mediator} \\
\midrule
\textbf{Support} & 1.00 & - & - & - \\
\textbf{Polarization} & 0.03 & 1.00 & - & - \\
\textbf{Community} & -0.01 & 0.00 & 1.00 & - \\
\textbf{Mediator} & 0.00 & 0.08 & 0.33 & 1.00 \\
\bottomrule
\end{tabular}
\caption{Correlation Coefficients for Independent Variables and Mediator (Model 1).}
\label{tab:cm_1}
\end{table*}

\begin{table*}
\small
\centering
\begin{tabular}{p{2.8cm}p{9cm}p{1cm}}
\toprule
\textbf{Category} & \textbf{Subcategory} &  \textbf{Count} \\
\midrule
Politics & General (27); Military (27); Economics (18); Gun Control (7) &   79 \\
Social justice & Immigration (27); LGBTQ+ (27); Abortion (21); Race (13); Diversity (5); Gender (5)  &    98 \\
Health & COVID\&Vaccination (9); Mental Health (3); Other (2) &    14\\
Technology & AI (9); Other (4) &    13 \\
Religion & --- &    16 \\
Environment & --- &    14 \\
Popular Culture & Celebrities (8); Sports (2) &    10\\
Personal or Local & --- &    16 \\
None & No disparity in posts (137); No topics provided (12) &    149 \\
\bottomrule
\end{tabular}
\caption{Topics for which survey respondents report a disparity between their private beliefs and their social media posts (from open-ended questions).}
\label{tab: survey_topic}
\end{table*}

\begin{table*}[ht]
\centering
\setlength{\tabcolsep}{2.7pt} 
\renewcommand{\arraystretch}{1} 
\begin{tabular}{@{}lcccc@{}} 
\toprule
 & \textbf{Support} & \textbf{Polarization} & \textbf{Community} & \textbf{Mediator} \\
\midrule
\textbf{Support} & 1.00 & - & - & - \\
\textbf{Polarization} & 0.09 & 1.00 & - & - \\
\textbf{Community} & -0.02 & 0.03 & 1.00 & - \\
\textbf{Mediator} & 0.01 & 0.10 & 0.33 & 1.00 \\
\bottomrule
\end{tabular}
\caption{Correlation coefficients for independent variables and mediator (Model 2).}
\label{tab:cm_2}
\end{table*}

\begin{table*}[ht]
\centering
\begin{tabular}{lcc}
\toprule
\textbf{Variable} & \textbf{Model 1 VIF} & \textbf{Model 2 VIF} \\
\midrule
Support & 1.00 & 1.01 \\
Polarization & 1.01 & 1.02 \\
Community Size (maximum\_total) & 1.12 & 1.12 \\
Frequency (frequency\_numeric) & 1.13 & 1.13 \\
\bottomrule
\end{tabular}
\caption{Variance inflation factors (VIF) for independent variables and mediator in Model 1 and Model 2.}
\label{tab:vif_models}
\end{table*}

\begin{table*}[ht]
\centering
\begin{tabular}{@{}lccc@{}}
\toprule
\textbf{Path} & \textbf{Coefficient} & \textbf{Std. Err} & \textbf{CI} \\ \midrule
Community Size to Frequency of Posting &   0.33&0.06  &[0.29, 0.45]  \\
Frequency of Posting to Dependent Variable &-0.53  &0.23  &[-0.98, -0.07]\\
Direct Effect of Community Size  & 0.19 & 0.10  & [0.03, 0.38] \\
Indirect Effect  & -0.18& - & [-0.22, -0.04] \\ 
\bottomrule
\end{tabular}
\caption{Mediation analysis results: Effect of community size on decision to remain silent mediated by frequency of posting (The indirect effect coefficient is derived from the average of 1,000 bootstrap samples, with the confidence interval calculated via bootstrapping.)}
\label{tab:mediation_1}
\end{table*}

\begin{table*}[ht]
\centering
\begin{tabular}{@{}lccc@{}}
\toprule
\textbf{Path} & \textbf{Coefficient} & \textbf{Std. Err} & \textbf{CI} \\ \midrule
Community Size to Frequency of Posting  &  0.33 & 0.09 &[0.15,0.51]  \\
Frequency of Posting to Dependent Variable  & -0.10 &0.06 & [-0.22, 0.03]\\
Direct Effect of Community Size  & 0.04&  0.05 &[-0.06,0.13] \\
Indirect Effect  & -0.03 & - &[-0.07, -0.00]  \\ 
\bottomrule
\end{tabular}
\caption{Mediation analysis results: effect of community size on discrepancy between public and private opinion mediated by frequency of posting (The indirect effect coefficient is derived from the average of 1000 bootstrap samples, with the confidence interval calculated via bootstrapping.)}
\label{tab:mediation_2}
\end{table*}

\end{document}